\documentclass[aps,prb,a4paper,10pt,twocolumn,showpacs,floatfix,superscriptaddress,preprintnumbers,longbibliography]{revtex4-2}
\usepackage{float}
\usepackage{dcolumn,graphicx,color,booktabs,microtype,afterpage}
\usepackage{amssymb}
\usepackage{amsmath}
\usepackage[charter,greekuppercase=italicized]{mathdesign}
\usepackage{sidecap}
\usepackage[mathlines]{lineno}

\graphicspath{{./}{figure/}}
\renewcommand{\tablename}{Table}
\makeatletter\renewcommand{\fnum@figure}[1]{\figurename~\thefigure.~}\makeatother
\makeatletter\renewcommand{\fnum@table}[1]{\tablename~\thetable.}\makeatother

\newcount\hh \newcount\mm
\hh=\time \divide\hh by 60
\mm=\hh \multiply\mm by 60 \mm=-\mm
\advance\mm by \time
\def\now{\number\hh:\ifnum\mm<10{}0\fi\number\mm}

\usepackage[colorlinks,plainpages=false,linkcolor=blue,urlcolor=blue,citecolor=blue,pdfpagemode=UseNone,pdfstartview=FitBH]{hyperref}
\usepackage{nicefrac}

\newcommand{\tcr}[1]{\textcolor{black}{#1}}

\begin{document}

\makeatletter\renewcommand{\ps@plain}{%
\def\@evenhead{\hfill\itshape\rightmark}%
\def\@oddhead{\itshape\leftmark\hfill}%
\renewcommand{\@evenfoot}{\hfill\small{--~\thepage~--}\hfill}%
\renewcommand{\@oddfoot}{\hfill\small{--~\thepage~--}\hfill}%
}\makeatother\pagestyle{plain}

\preprint{\textit{Preprint: \today, \now}} %For internal use only, do not distribute.}}
%\linenumbers
%\title{Fully-gapped superconducting state with preserved time-reversal symmetry in noncentrosymmetric NbReSi  superconductor}
%\title{Centro- vs. non-centrosymmetric Re-B superconducting alloys: An $\mu$SR study}
\title{Multiband superconductivity in the topological Kramers nodal-line semimetals}
%\title{Evidence of spin-triplet pairing in a fully-gapped noncentrosymmetric superconductor}

%
\author{Tian Shang}\email[Corresponding authors:\\]{tshang@phy.ecnu.edu.cn}
\affiliation{Key Laboratory of Polar Materials and Devices (MOE), School of Physics and Electronic Science, East China Normal University, Shanghai 200241, China}
%\affiliation{Laboratory for Multiscale Materials Experiments, Paul Scherrer Institut, Villigen CH-5232, Switzerland}
%
%%
\author{Jianzhou Zhao}\email[Corresponding authors:\\]{jzzhao@swust.edu.cn}
\affiliation{Co-Innovation Center for New Energetic Materials, Southwest University of Science and Technology, Mianyang, 621010, China} 
\author{Keqi Xia}
\affiliation{Key Laboratory of Polar Materials and Devices (MOE), School of Physics and Electronic Science, East China Normal University} %\affiliation{Laboratorium f\"ur Festk\"orperphysik, ETH Z\"urich, CH-8093 Z\"urich, Switzerland}
\author{Lun-Hui Hu}
\affiliation{Center for Correlated Matter and Department of Physics, Zhejiang University, Hangzhou, 310058, People’s Republic of China}
\author{Yang Xu}
\affiliation{Key Laboratory of Polar Materials and Devices (MOE), School of Physics and Electronic Science, East China Normal University, Shanghai 200241, China}
\author{Qingfeng Zhan}
\affiliation{Key Laboratory of Polar Materials and Devices (MOE), School of Physics and Electronic Science, East China Normal University, Shanghai 200241, China}

\author{Dariusz Jakub Gawryluk}
%\affiliation{Laboratory for Multiscale Materials Experiments, Paul Scherrer Institut, CH-5232 Villigen PSI, Switzerland}
\affiliation{PSI Center for Neutron and Muon Sciences CNM, 5232 Villigen PSI, Switzerland}
\author{Toni Shiroka}\email[Corresponding authors:\\]{tshiroka@phys.ethz.ch}
\affiliation{PSI Center for Neutron and Muon Sciences CNM, 5232 Villigen PSI, Switzerland}
\affiliation{Laboratorium f\"ur Festk\"orperphysik, ETH Z\"urich, CH-8093 Z\"urich, Switzerland}
%\affiliation{Laboratory for Muon-Spin Spectroscopy, Paul Scherrer Institut, Villigen PSI, Switzerland}
%
%
\begin{abstract}
%The ruthenium-based ternary silicides, which have been classified as three-dimensional Kramers nodal-line semimetals, are characterized by large antisymmetric spin-orbit couplings and hourglass-like dispersions near the Fermi level. 
Recent band-structure calculations predict that the ruthenium-based ternary silicides are three-dimensional Kramers nodal line semimetals. 
%\tcr{The ruthenium-based ternary silicides \tcb{are predicted to be} three-dimensional Kramers nodal line semimetals \tcb{by recent} band-structure calculations.} 
Among them, NbRuSi and TaRuSi show bulk superconductivity (SC) below $T_c \sim 3$\,K and 4\,K, as well as spontaneous magnetic fields. The latter indicates the breaking of time-reversal symmetry and, thus, unconventional SC in both compounds. Previous temperature-dependent muon-spin spectroscopy studies failed to distinguish whether such compounds exhibit single-gap or multi-gap SC. Here, we report on systematic measurements of the field-dependent muon-spin relaxation rates in the superconducting state and on temperature-dependent electrical resistivity and specific heat under applied magnetic fields. Both the upper critical field and the field-dependent superconducting relaxation are well described by a two-band model. By combining our experimental results with numerical band-structure calculations, we provide solid evidence for multiband SC in NbRuSi and TaRuSi, and thus offer further insight into the unconventional- and topological nature of their superconductivity.
\end{abstract}

\maketitle\enlargethispage{3pt}

\vspace{-5pt}
\section{\label{sec:Introduction}Introduction}\enlargethispage{8pt}
Symmetry, including its space-translation/rotation, time-reversal, and 
and parity variants plays a significant role in determining the physical properties of solid-state systems. A combination of selected symmetry elements often leads to exotic
quasiparticle excitations, analogous to the particles predicted in high-energy physics,
such as, Dirac-, Weyl-, or Majorana fermions~\cite{Armitage2018,Lv2021,Yan2017,Wieder2022,Bradlyn2017,Vergniory2019,Vergniory2019,Tang2019,Bernevig2022}. Materials that lack an inversion center are among the best candidates
for exploring the resulting topological phenomena. 
For instance, Weyl~\cite{Xu2015a,Xu2015b,Lv2015,Xu2016,Souma2016}, hourglass~\cite{Wang2016,Wu2019,Shang2022,Shang2023}, Kramers nodal-line ~\cite{Xie2021,Shang2022,Shang2023,Chang2018}, and multifold chiral fermions~\cite{Chang2018,Bradlyn2016,Rao2019,Sanchez2019} have been
experimentally observed or predicted to occur in different
families of noncentrosymmetric materials. 

To date, research on topological materials has been primarily focused on the case of noninteracting
electronic bands. On the contrary, the interplay between
topology and correlated electronic states, such as superconductivity
(SC) or magnetism, remains less explored.
In addition to the topological features, the noncentrosymmetric materials that exhibit SC, known as noncentrosymmetric superconductors (NCSCs), represent one of the ideal platforms to study such interplay and  
to search for unconventional and topological SC, as well as Majorana zero modes, with potential applications to quantum computation~\cite{Sato2017,Qi2011,Kallin2016}. When the bulk of the material transitions into the superconducting state,
the proximity effect can lead to topological superconducting
surface states. Such topologically protected surface states have been
proposed to occur, for instance, in the noncentrosymmetric $\beta$-Bi$_2$Pd and
PbTaSe$_2$ superconductors~\cite{Guan2016,Sakano2015}.
In addition, NCSCs often exhibit a variety of exotic  superconducting properties due to the occurrence of 
admixtures of spin-singlet and spin-triplet pairing, e.g., upper critical fields beyond the Pauli 
limit~\cite{Carnicom2018,Bauer2004,Su2021a}, nodes in the superconducting gap~\cite{yuan2006,nishiyama2007,bonalde2005CePt3Si,Shang2020}, 
multigap SC~\cite{kuroiwa2008,Sundar2021}, or a breaking of
time-reversal symmetry (TRS) in the superconducting 
state~\cite{Shang2020,Hillier2009,Barker2015,Shang2020b,Singh2014,Shang2018b,Shang2022}.

\begin{figure}[!htp]
	\centering
	\vspace{-1ex}%
	\includegraphics[width=0.48\textwidth,angle=0]{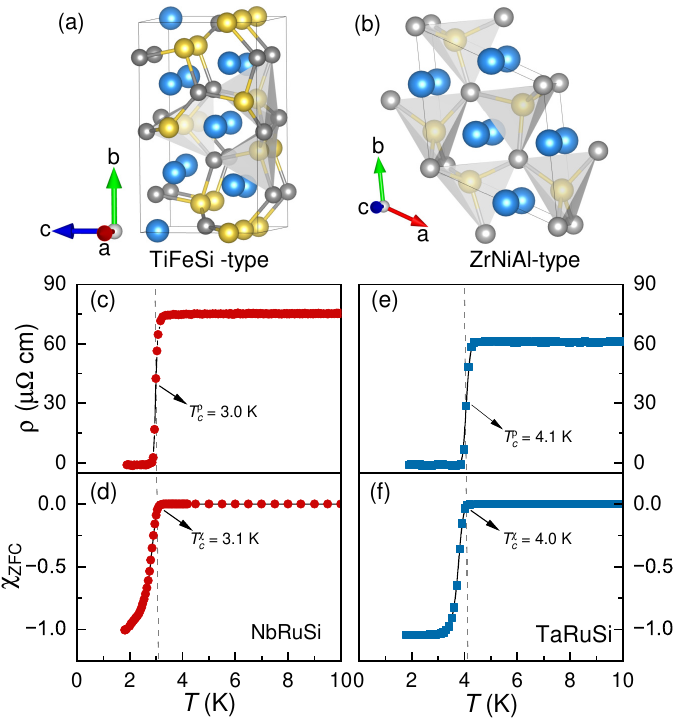}
	\caption{\label{fig:structure} Crystal structure (unit cell) of orthorhombic TiFeSi-type and hexagonal ZrNiAl-type materials. Blue, yellow, and gray spheres stand for the Ti(Zr), Fe(Ni), and Si(Al) atoms, respectively. Both structures lack an inversion center. Temperature-dependent electrical resistivity $\rho(T)$ collected
	using a $^{4}$He refrigerator (c) and magnetic susceptibility $\chi_\mathrm{ZFC}(T)$ (d) for NbRuSi.
		The analog results for TaRuSi are shown in panels (e) and (f). Magnetic-susceptibility data (taken from Ref.~\cite{Shang2022}) were collected in a field of 1\,mT, applied after zero-field cooling (ZFC). The dashed lines mark the $T_c$ values  determined from $\rho(T)$ and $\chi_\mathrm{ZFC}(T)$. Both are highly consistent with the values determined from heat-capacity and {\textmu}SR measurements (see below).}
\end{figure}

The ternary transition-metal silicides, germanides, and pnictides adopt a few distinct 
crystal structures, including the tetragonal PbClF-type ($P$4/$nmmZ$, No.~129)~\cite{Welter1993}, orthorhombic TiNiSi-type ($Pnma$, No.~62)~\cite{Morozkin1999}, hexagonal ZrNiAl-type ($P\overline{6}2m$, No.~189)~\cite{Subba1985}, and orthorhombic TiFeSi-type ($Ima2$, No.~46)~\cite{Subba1985}.
Among the four crystal structures, the ZrNiAl and TiFeSi types belong
to the noncentrosymmetric class that lacks an inversion center [see Fig.~\Ref{fig:structure}(a)-(b)]. Materials with these two structures have often been investigated
in search of new exotic electronic properties.

%==== figure =============================%
%=== end figure ==========================%

Several ZrNiAl-type compounds are known to exhibit SC~\cite{Meisner1983,Ichimin1999,Su2021a}.
Most of them display an ordinary fully-gapped superconducting state with preserved TRS, hence, are conventional superconductors~\cite{Das2021,Das2023,Su2021a}. 
Yet, NbReSi shows a large upper critical field, beyond the weak-coupling Pauli limiting field (i.e., $H_\mathrm{P}$ = 1.86$T_c$), which hints at the presence of an unconventional pairing~\cite{Shang2022b}. 
Likewise, the ZrNiAl-type HfRhGe represents another notable exception, since its
fully-gapped SC was recently found to break the TRS~\cite{Sajilesh2024}. In addition, 
%multiple Weyl nodes near the Fermi level and surface Fermi arcs
%dispersing across the Fermi level were suggested to occur in this material~\cite{Sajilesh2024}.
%\tcr{multiple Weyl nodes near the Fermi level and surface Fermi arcs dispersing across the Fermi level were suggested from band structure calculations in this material~\cite{Sajilesh2024}.}
band-structure calculations on HfRhGe suggested the presence
of surface Fermi arcs and of multiple Weyl nodes near the
Fermi level~\cite{Sajilesh2024}.
% This part below seems to me redundant, with too many repetitions of "Fermi level". TS 
%dispersing across the Fermi level~\cite{Sajilesh2024}.} 
As for the TiFeSi-type materials, such as $T$RuSi ($T$ = Ti, Nb,
Hf, Ta) or TaReSi, band-structure calculations predict that their
normal states are three-dimensional Kramers nodal-line semimetals,
characterized by a large antisymmetric spin-orbit coupling and by
hourglass-like dispersions near the  Fermi level~\cite{Shang2022,Shang2023}.
While TaReSi behaves as a conventional superconductor, both
NbRuSi and TaRuSi spontaneously break the TRS in the superconducting state and adopt a unitary ($s + ip$) pairing~\cite{Shang2022},
corresponding to a mixture of singlet- and triplet pairing.
The origin of the differing superconducting properties of the ZrNiAl- and TiFeSi-type NCSCs
are not yet understood and clearly deserve
further investigation. In any case, 
the presence of nontrivial electronic bands near the Fermi level 
suggests them as a promising platform for investigating topological SC.

Among the above mentioned NCSCs, some of them show multiband (or multigap)
SC~\cite{Shang2022b}. In the case of NbRuSi and TaRuSi, although 
numerical band-structure calculations reveal that multiple electronic bands %are identified to
cross the Fermi level, our previous temperature-dependent muon-spin relaxation and rotation ({\textmu}SR) measurements in both compounds failed to distinguish a single-gap from a multigap SC.
This could be due to either the comparable gap sizes or to the small weight of the second gap~\cite{Shang2022}. In this paper, by performing new field-dependent {\textmu}SR measurements in the superconducting state, as well as temperature-dependent electrical-resistivity and heat-capacity measurements under various magnetic fields, we provide solid evidence for the multiband
nature of superconductivity in both NbRuSi and TaRuSi.

\section{Experimental details\label{sec:details}}\enlargethispage{8pt}

Polycrystalline NbRuSi and TaRuSi samples were prepared by the arc-melting method. Full details of the synthesis and the structural characterization were previously reported in Ref.~\cite{Shang2022}. 
The bulk SC was characterized
by elec\-tric\-al\--re\-sis\-ti\-vi\-\mbox{ty-,} heat-ca\-pac\-i\-\mbox{ty-,} and magnetization measurements, which were performed on a Quantum Design magnetic property measurement system and a physical property measurement system.
The upper critical fields $H_\mathrm{c2}$ were determined by the above measurements performed under various magnetic fields up to 5\,T. 
Bulk {\textmu}SR measurements were carried out at the multipurpose 
surface-muon spectrometer (Dolly) of the Swiss muon source at Paul 
Scherrer Institut, Villigen, Switzerland. 
In this study, in order to investigate the possible
multiband SC of NbRuSi and TaRuSi, we performed transverse-field (TF)-{\textmu}SR measurements in various magnetic fields up to 0.8\,T at the base temperature (0.3\,K). The samples were mounted on a
25-{\textmu}m-thick copper foil which, while ensuring thermalization at low
temperatures, was thin enough to provide essentially background-free data. The time-differential {\textmu}SR spectra were collected after a field-cooling protocol and then analyzed by means of the \texttt{musrfit} software package~\cite{Suter2012}.

First-principles calculations were performed based on the density functional theory (DFT), as implemented in the \texttt{Quantum ESPRESSO} package~\cite{giannozzi2009,giannozzi2017}.
The exchange-correlation functional was treated with the generalized gradient approximation using the Perdew-Burke-Ernzerhof (PBE) realization~\cite{Perdew1996iq}.
The projector augmented wave pseudopotentials were adopted~\cite{Blochl1994zz}.
We considered 13 electrons for Nb (4$s^{2}$5$s^{2}$4$p^6$4$d^{3}$),
13 electrons for Ta (5$s^{2}$6$s^{2}$5$p^6$5$d^{3}$), 16 electrons for Ru
(4$s^{2}$5$s^{2}$4$p^6$4$d^{6}$) and 4 electrons for Si (3$s^{2}$3$p^{2}$)
as valence electrons.
Spin-orbit coupling effects were included in the calculation.
The kinetic energy cutoff for the wavefunctions was set to 60 Ry, 
while for the charge density it was fixed to 600 Ry. 
For the self-consistent calculations, the Brillouin zone integration 
was performed on a Monkhorst-Pack grid mesh of $12 \times 12 \times 12$ $k$-points.
The convergence criterion was set to $10^{-7}$ Ry. 

%=== begin figure ==========================%
\begin{figure*}[!htp]
	\centering
	%	\vspace{-1ex}%
	\includegraphics[width=0.9\textwidth,angle=0]{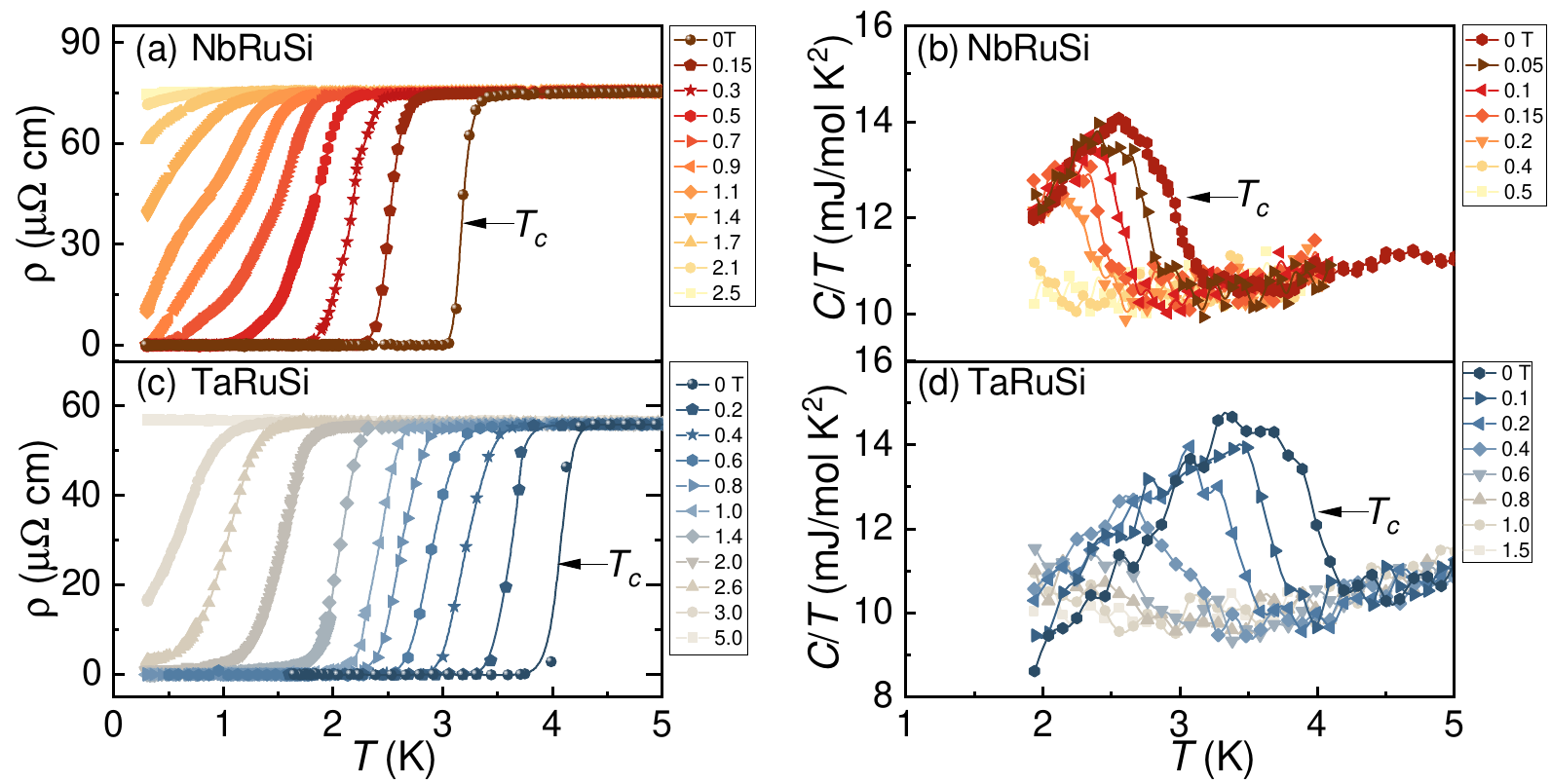}
	\caption{\label{fig:rho} 
		 NbRuSi temperature-dependent electrical resistivity
		$\rho(T,H)$ (a), measured using a $^{3}$He refrigerator,
		and specific heat $C(T,H)/T$ (b), measured
		at various magnetic fields. For both datasets, the $T_c$ values were defined as the midpoint of the superconducting transition, shown here with an arrow for the zero-field case. The analog results for TaRuSi are shown in panels (c) and (d). 
		%\tcr{Here, the resistivity data were collected using a He3 refrigerator}.
		}
\end{figure*}
%=== end figure ==========================%

\section{Results and discussion\label{sec:results}}\enlargethispage{8pt}
%\subsection{$\mu$SR study}
%
The temperature-dependent electrical resistivity $\rho(T)$ and magnetic susceptibility $\chi_\mathrm{ZFC}(T)$ data in the low-$T$ region are plotted in Fig.~\ref{fig:structure}(c)-(d) and Fig.~\ref{fig:structure}(e)-(f) for NbRuSi and TaRuSi, respectively. 
Clearly, both $\rho(T)$ and $\chi_\mathrm{ZFC}(T)$ exhibit sharp superconducting transitions.  
As shown by the dashed lines, in the NbRuSi case, the $T_c = 3.0$\,K determined from the midpoint of the superconducting transition in $\rho(T)$ is consistent with the onset transition temperature $T_c = 3.1$\,K in $\chi_\mathrm{ZFC}(T)$. In the TaRuSi case, $T_c = 4.1$ and 4.0\,K were identified from $\rho(T)$ and $\chi_\mathrm{ZFC}(T)$, respectively. This consistency allowed us to use the $T_c$ values determined from
the resistivity $\rho(T)$ to evaluate the upper critical fields of both compounds.
%
%=== begin figure ==========================%
\begin{figure}[!htp]
	\centering
	\includegraphics[width=0.46\textwidth,angle= 0]{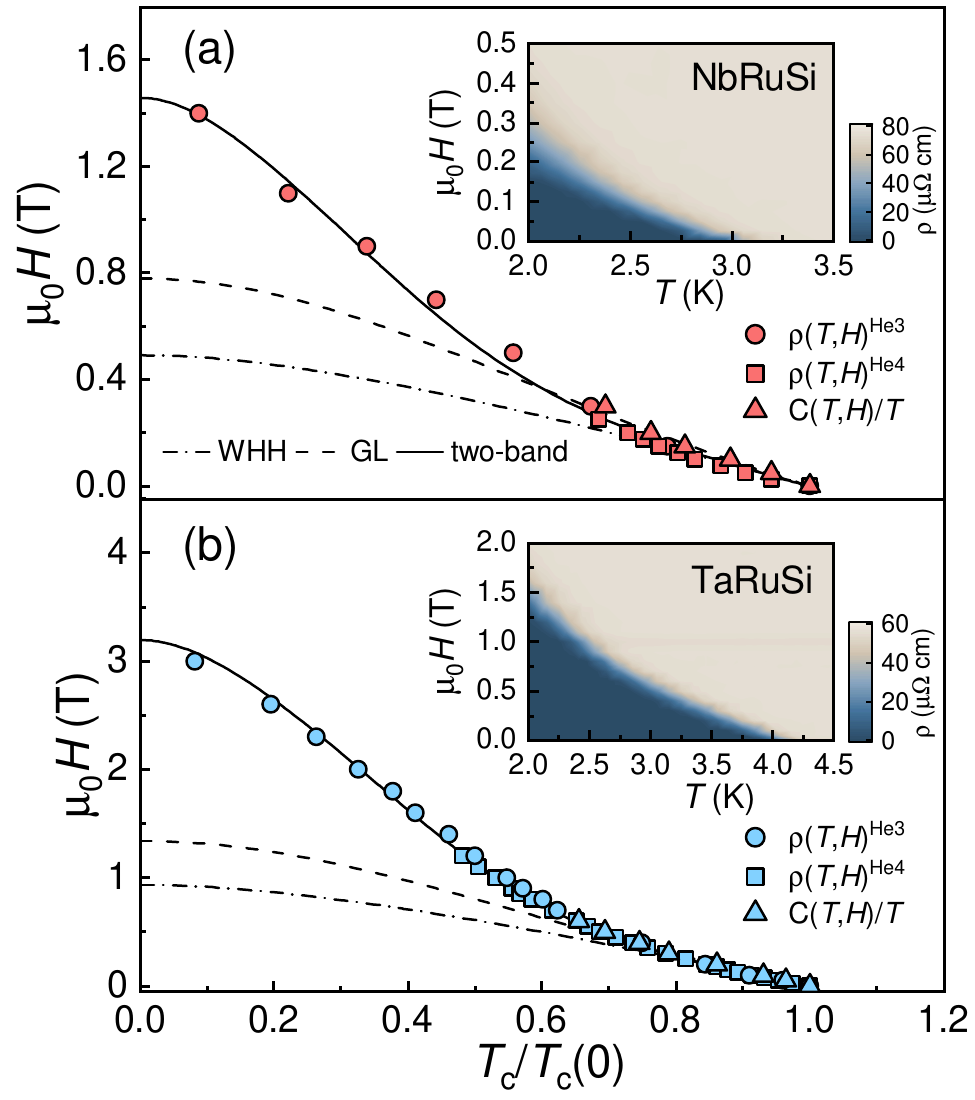}
	\caption{\label{fig:Hc2}%
	Superconducting transition temperature $T_c$ vs applied magnetic
	field for NbRuSi (a) and TaRuSi (b), as determined from
	temperature-dependent electrical-resistivity $\rho(T,H)$ and
	specific-heat $C(T,H)/T$ measurements (see Fig.~\ref{fig:rho}).
    All $T_c$ values were normalized to the zero-field value $T_c(0)$.
	The contour plots of $\rho(T,H)$ in the insets clearly show a
	positive curvature for both NbRuSi and TaRuSi.
		%\tcb{The $\rho(T,H)$ data were collected using a \tcb{$^{4}$He} refrigerator.}
	Three different fit models, including the WHH (dash-dotted line),
	the GL (dashed line), and the two-band model (solid line), were used
	to analyze the $H_\mathrm{c2}(T)$ data.}
\end{figure}
%=== end figure ==========================%  

To obtain the upper critical fields $H_\mathrm{c2}$ of NbRuSi and Ta\-Ru\-Si, we measured the temperature-dependent electrical 
resistivity $\rho(T,H)$ and specific heat $C(T,H)/T$ at various magnetic fields up to 5\,T. As shown in Fig.~\ref{fig:rho}, upon increasing the magnetic field, the superconducting transition in both $\rho(T)$ and $C(T)/T$ shifts to lower temperatures, and becomes broader. 
Figure~\ref{fig:Hc2} summarizes the reduced superconducting transition 
temperature $T_c$/$T_c(0)$ versus the magnetic field, where $T_c(0)$ is the zero-field superconducting transition temperature.  
The $T_c$ values, as determined from the midpoint of the superconducting transition in both $\rho(T)$ and $C(T)/T$ (see arrows in Fig.~\ref{fig:rho})
%as determined using different techniques, 
are highly consistent.  
The $H_\mathrm{c2}(T)$ data were analyzed by means of the Ginzburg-Landau (GL)~\cite{Zhu2008}, Werthamer-Helfand-Hohenberg (WHH)~\cite{Werthamer1966}, and two-band models~\cite{Gurevich2011}. As shown in Fig.~\ref{fig:Hc2}, the GL- and WHH models describe reasonably well the low-field $H_\mathrm{c2}(T)$ data, i.e., $\mu_0H <$ 0.3\,T (0.6\,T) of NbRuSi (TaRuSi). However, at higher magnetic fields, both models deviate significantly from the experimental data and yield underestimated zero-temperature upper critical fields $H_\mathrm{c2}(0)$. 
Such a discrepancy is already an indication of multiband SC in NbRuSi and TaRuSi, further reinforced by the positive curvature of $H_\mathrm{c2}(T)$, a typical feature of multiband superconductors, such as Lu$_2$Fe$_3$Si$_5$~\cite{Nakajima2012}, MgB$_2$~\cite{Muller2001,Gurevich2004}, 
and ReB binary alloys~\cite{Shang2021}. 
A positive curvature in $H_\mathrm{c2}(T)$ usually reflects the suppression of the smaller superconducting gap as the magnetic field increases. As clearly demonstrated in the insets of Fig.~\ref{fig:Hc2}, the contour plots of $\rho(T,H)$ exhibit clear positive curvatures for both NbRuSi and TaRuSi.  
As a consequence, only the two-band model 
shows a remarkable agreement with the experimental data in the full temperature- or field range and it provides 
$\mu_0 H_\mathrm{c2}(0)$ = 1.40(5)\,T and 3.20(5)\,T for NbRuSi and TaRuSi, respectively (see solid lines in Fig.~\ref{fig:Hc2}). Both  $H_\mathrm{c2}(0)$ values are much smaller than the Pauli-limiting field (i.e., $H_\mathrm{P}$ $\sim$ 1.86$T_c$). Thus, the orbital pair-breaking effect is dominant in both NbRuSi and TaRuSi. Note that, to study
their multigap nature further, heat-capacity measurements down to lower
temperatures ($\sim 0.3$\,K) remain crucial.

%=== begin figure ==========================%
\begin{figure*}[!htp]
	\centering
	\includegraphics[width=0.9\textwidth,angle= 0]{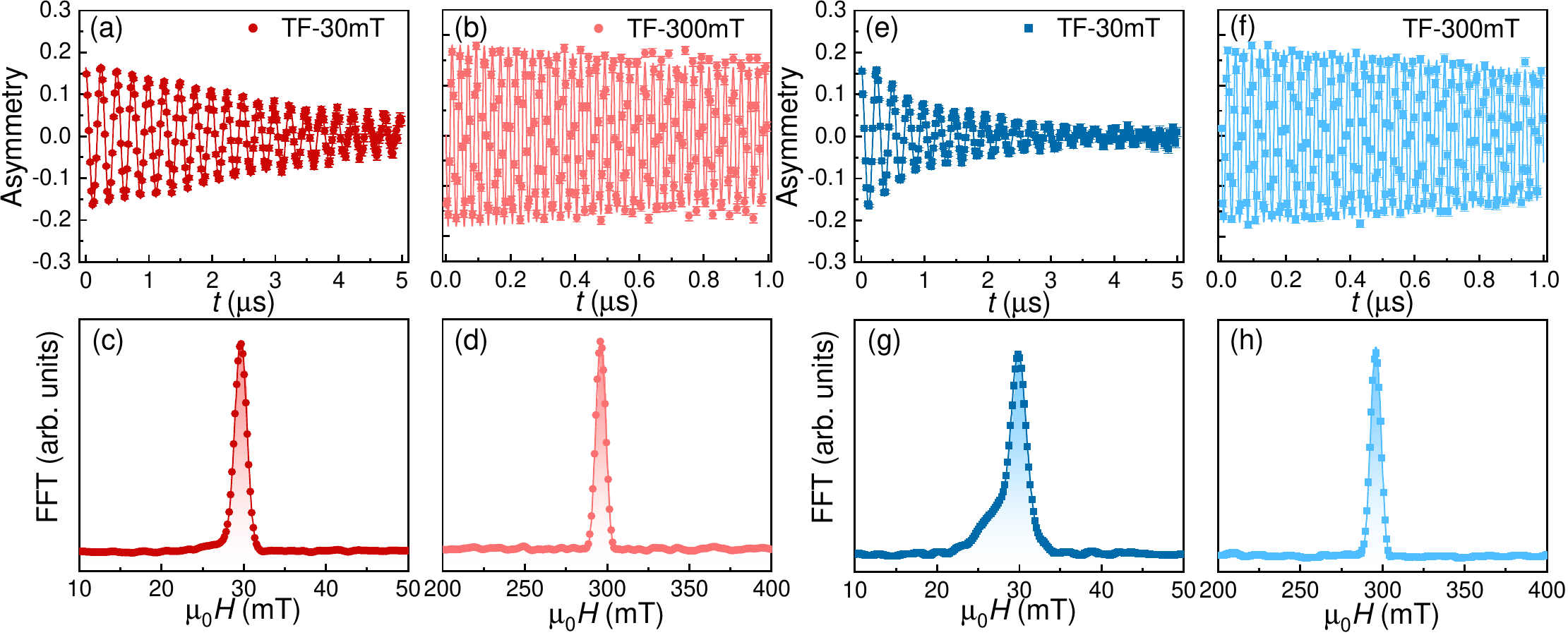}
	\caption{\label{fig:muSR}
		TF-{\textmu}SR spectra of NbRuSi measured at $T = 0.3$\,K (superconducting state) in a field of 30\,mT (a) and 300\,mT (b). Each spectrum was collected using a field-cooling protocol. The fast Fourier transforms of the
		relevant {\textmu}SR spectra are shown in panels (c) and (d). 
		The analog results for TaRuSi are shown in panels (e)-(h).
		The solid lines through the data are fits to Eq.~\eqref{eq:TF_muSR} using two oscillations.}
\end{figure*}
%=== end figure ==========================% 
%

To gain further insight into the multiband SC of NbRuSi and TaRuSi,
we performed TF-{\textmu}SR measurements at various applied
magnetic fields up to 0.8\,T at base temperature (0.3\,K). 
In fact, our previous temperature-dependent TF-{\textmu}SR measurements
failed to distinguish a single-gap from a multiple-gap scenario
in either compound~\cite{Shang2022}.
We attribute this to the comparable
gap sizes or to the small weight of the additional gap. 
To track the additional field-distribution broadening due to the flux-line lattice (FLL) in the mixed superconducting 
state, a magnetic field  was applied in the normal state and then the sample was cooled down to base temperature (0.3\,K),  
where the {\textmu}SR spectra were collected. Representative TF-{\textmu}SR spectra collected in a field of 30\,mT and 300\,mT at base temperature are shown in Fig.~\ref{fig:muSR}(a)-(b) and Fig.~\ref{fig:muSR}(e)-(f) for NbRuSi and TaRuSi, respectively.
As the magnetic field increases, the muon-spin relaxation rate decreases. As a consequence, the field-distribution broadening due to the FLL also decreases, which is clearly reflected in the fast-Fourier transform (FFT) spectra of the corresponding TF-{\textmu}SR spectra in Fig.~\ref{fig:muSR}(c)-(d) and Fig.~\ref{fig:muSR}(g)-(h) for NbRuSi and TaRuSi, respectively.  Similar to the temperature-dependent measurements~\cite{Shang2022}, the TF-{\textmu}SR spectra were also modeled by (see the solid lines in Fig.~\ref{fig:muSR}):  
\begin{equation}
	\label{eq:TF_muSR}
	A(t) = \sum\limits_{i=1}^2 A_i \cos(\gamma_{\mu} B_i t + \phi) e^{- \sigma_i^2 t^2/2} +
	A_\mathrm{bg} \cos(\gamma_{\mu} B_\mathrm{bg} t + \phi).
\end{equation}
Here $A_{i}$, $A_\mathrm{bg}$ and $B_{i}$, $B_\mathrm{bg}$ 
are the asymmetries and the local fields sensed by the implanted
muons in the sample and copper sample holder, 
$\gamma_{\mu}$/2$\pi = 135.53$\,MHz/T 
is the muon gyromagnetic ratio, $\phi$ is a shared initial phase, and $\sigma_{i}$ 
is the Gaussian relaxation rate of the $i$th component. 
The effective Gaussian relaxation rate can be estimated from 
$\sigma_\mathrm{eff}^2/\gamma_\mu^2 = \sum_{i=1}^2 A_i [\sigma_i^2/\gamma_{\mu}^2 + \left(B_i - \langle B \rangle\right)^2]/A_\mathrm{s}$, where $\langle B \rangle = (A_1\,B_1 + A_2\,B_2)/A_\mathrm{s}$ and $A_\mathrm{s} = A_1 + A_2$~\cite{Maisuradze2009}. Then, 
the superconducting contribution to the muon-spin relaxation rate can be
extracted from $\sigma_\mathrm{sc} = \sqrt{\sigma_\mathrm{eff}^{2} - \sigma^{2}_\mathrm{n}}$, where $\sigma_\mathrm{n}$ is the nuclear relaxation rate in the normal state.

\begin{figure}[!htp]
	\centering
	\includegraphics[width=0.49\textwidth,angle=0]{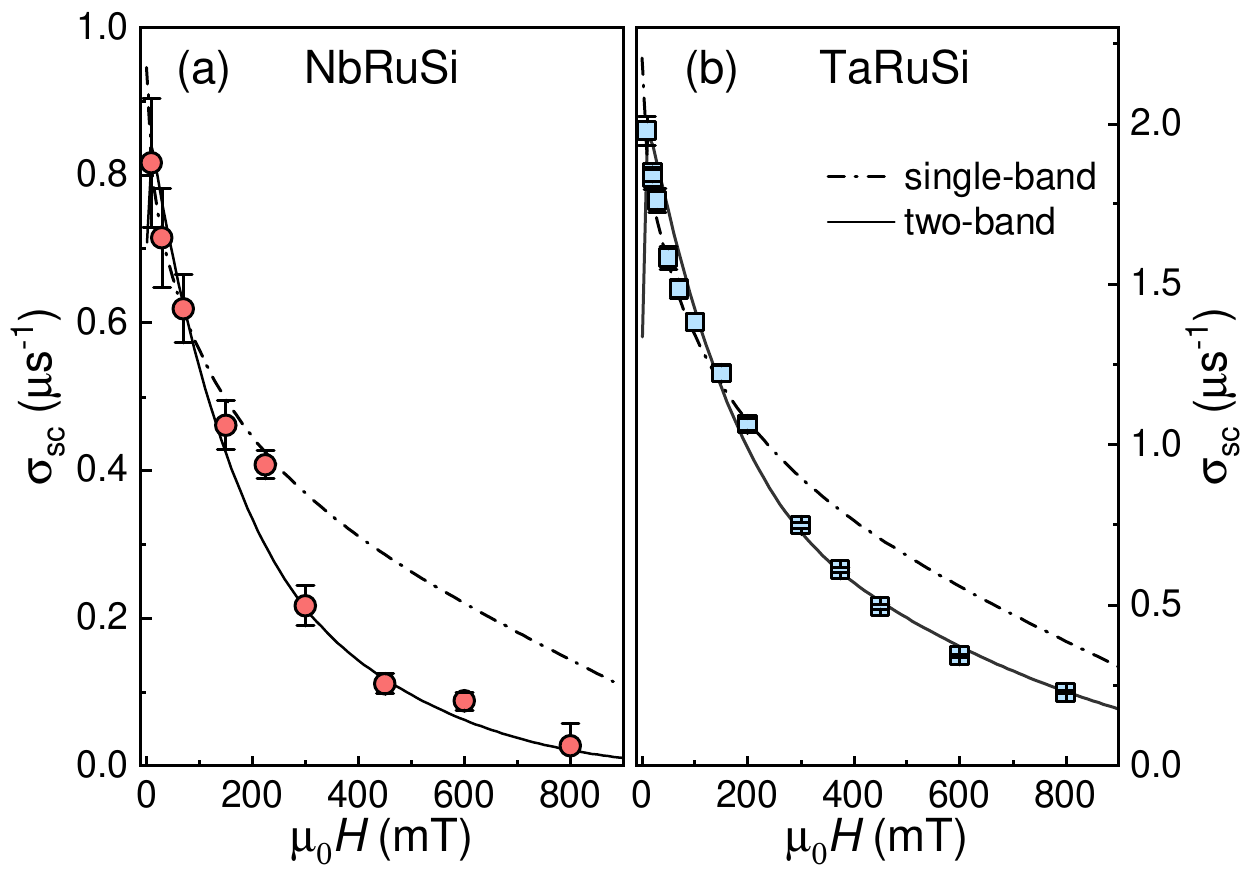}
	\vspace{-2ex}%
	\caption{\label{fig:relaxation}
		Field-dependent superconducting Gaussian relaxation rate $\sigma_\mathrm{sc}(H)$
		at $T = 0.3$\,K for NbRuSi (a) and TaRuSi (b). The dash-dotted and solid lines represent fits to the single-band
		and the two-band model, respectively.}
		 %\tcr{To avoid the suspicious effects at $H \le H_\mathrm{c1}$, here the field-dependent {\textmu}SR spectra were collected at {\textmu}$_0$$H$ $\ge$ 10\,mT}.}
\end{figure}

The obtained superconducting Gaussian relaxation rates $\sigma_\mathrm{sc}$
(at $T = 0.3$\,K) as a function of the applied magnetic field are
summarized in Fig.~\ref{fig:relaxation}. Both NbRuSi and TaRuSi show
a similar relaxation rate $\sigma_\mathrm{sc}(H)$,
which decreases significantly
as the magnetic field increases. 
We used both a singe-band and a two-band model to analyze
the $\sigma_\mathrm{sc}(H)$ data. In the case of a single-band or a single-gap superconductor, $\sigma_\mathrm{sc}(H)$ generally 
follows the relation~\cite{Barford1988,Brandt2003}:
%%%%%%%%%%%%%%%%%%%%%%%%
\begin{equation}
	\label{eq:single-band}
	\sigma_\mathrm{sc} = 0.172 \frac{\gamma_{\mu} \Phi_0}{2\pi}(1-h)[1+1.21(1-\sqrt{h})^3]\lambda_0^{-2},
\end{equation}
%%%%%%%%%%%%%%%%%%%%%%
where $h = H_\mathrm{ex}/H_\mathrm{c2}$, with $H_\mathrm{ex}$ being 
the externally applied magnetic field and $\lambda_0$ the magnetic
penetration depth. As shown by dash-dotted lines in
Fig.~\ref{fig:relaxation}, for fields above 200\,mT, the
single-band model shows a very poor agreement with the experimental data, 
yielding underestimated upper critical fields {\textmu}$_0$$H_\mathrm{c2} = 1.1(1)$
and 1.3(1)\,T at $T = 0.3$\,K for NbRuSi and TaRuSi, respectively.

%==== figure =============================%
%=== end figure ==========================%

%==== figure =============================%
\begin{figure}[!htp]
	\centering
	\includegraphics[width=0.48\textwidth,angle=0]{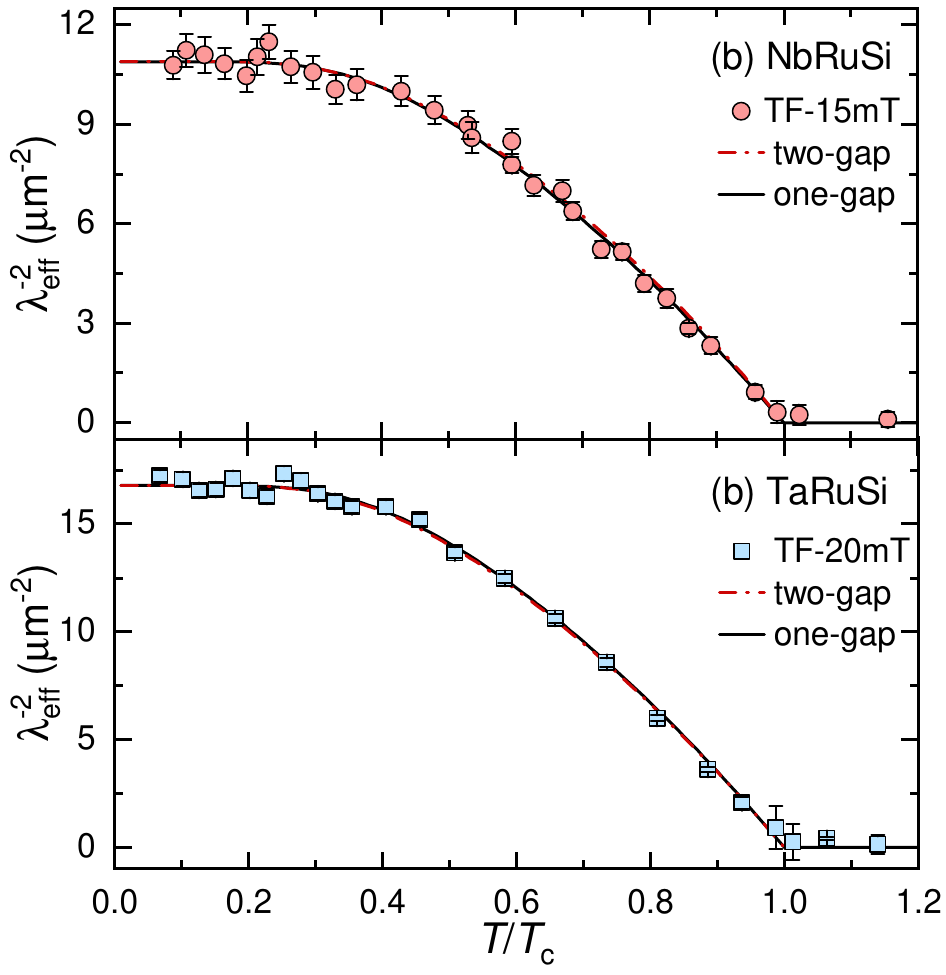}
	\vspace{-2ex}%
	\caption{\label{fig:lam}%
		Temperature-dependent inverse square of the effective magnetic penetration
		depth $\lambda_\mathrm{eff}^{-2}(T)$ for NbRuSi (a) and
		TaRuSi (b), as determined from TF-{\textmu}SR
		measurements in an applied magnetic field of 15\,mT and 20\,mT,
		respectively. The solid and dash-dotted lines represent fits to the $s$-wave model using one gap and two gaps, respectively. 
		Data of $\lambda_\mathrm{eff}^{-2}(T)$ were taken from Ref.~\cite{Shang2022}.
	}
\end{figure}
%=== end figure ==========================%

In a two-band model, each band is characterized by its own coherence 
length $\xi$ and a weight $w$ [or ($1-w$)], accounting for the relative contribution of each band to the total $\sigma_\mathrm{sc}$ and, hence, to the superfluid 
density~\cite{Serventi2004,Khasanov2014}.
In this case, the second moment of the field distribution
that determines $\sigma_\mathrm{sc}$ 
can be calculated in the framework of the modified London model
by means of the equation:
\begin{equation}
	\label{eq:modified_London}
	\langle B^2 \rangle = \frac{\sigma_\mathrm{sc}^2}{\gamma_\mu^2} = B^2 \sum\limits_{q \neq 0} \left[ w \frac{e^{-\frac{q^2\xi_1^2}{2(1-h_1)}}}{1+\frac{q^2\lambda_0^2}{1-h_1}} + (1-w) \frac{-e^{\frac{q^2\xi_2^2}{2(1-h_2)}}}{1+\frac{q^2\lambda_0^2}{1-h_2}} \right]^2.
\end{equation}
Here, $q = 4\pi/\sqrt{3}a (m\sqrt{3}/2, n + m/2)$ are the reciprocal 
lattice vectors for a hexagonal FLL, where $a$ is the inter-vortex distance, $m$ and $n$ are integer numbers. The mean field within the FLL is $B = {\mu}_0H$ for $H \gg  H_\mathrm{c1}$, where $H_\mathrm{c1}$ is the lower critical field;
$h_{1(2)} = H/H_\mathrm{c2,1(2)}$ is the reduced field within band
1(2) [the same as $h$ in Eq.~\eqref{eq:single-band}]
and $\xi_{1(2)}$ is the coherence length for the band 1(2).
As shown by the solid lines in Fig.~\ref{fig:relaxation}, the two-band
model is clearly superior to the single-band model. Hence, it
describes the $\sigma_\mathrm{sc}(H)$ data extremely well over the full field range. For NbRuSi, with $w$ = 0.7, the best fit gives $\lambda_0$ = 338(2)\,nm, $\xi_{1}$ = 19(1)\,nm, and $\xi_{2}$ = 13(1)\,nm; while for TaRuSi one obtains
$\lambda_0 = 216(2)$\,nm, $\xi_{1} = 18(1)$\,nm, and $\xi_{2} = 10.5(5)$\,nm, 
when using the same weight $w$ as for NbRuSi. 
The upper critical fields, calculated from the shortest coherence lengths $\xi_{2}$, $\mu_0H_\mathrm{c2} = \Phi_0/(2\pi\xi_2^{2})$, are 
1.9(3) and 3.0(3)\,T for NbRuSi and TaRuSi, comparable to the values determined from bulk measurements.
At the same time, by using the coherence length of the first band
$\xi_1$, the \emph{virtual} upper critical fields $\mu_0H_\mathrm{c2}^\ast$
are 0.9(1) and 0.8(1)\,T for NbRuSi and TaRuSi, respectively.
Most likely they correspond to the critical field which suppresses the smaller superconducting gap. The derived magnetic penetration
depths $\lambda_0$ are also consistent with the values extracted from the  temperature-dependent {\textmu}SR studies~\cite{Shang2022}. 
Clearly, the behavior of the field-dependent $\sigma_\mathrm{sc}(H)$ 
and the upper critical field $H_\mathrm{c2}(T)$ suggest a multiband
(or multigap) superconductivity in both NbRuSi and TaRuSi. \tcr{Noted that, 
	to avoid adverse effects at $H \le H_\mathrm{c1}$, the field-dependent {\textmu}SR spectra were collected at {\textmu}$_0$$H$ $\ge$ 10\,mT. A drop in the field-dependent relaxation could potentially be observed if the magnetic field were decrease below 10\,mT (see solid lines in Fig.~\ref{fig:relaxation}). However, this possibility was not investigated further. }

%==== Table =============================%
\begin{table}[!th]
	\centering
	\caption{Normal- and superconducting state properties of NbRuSi and
		TaRuSi, as determined from electrical-resistivity, magnetization,
		specific-heat, and {\textmu}SR measurements, as well as electronic
		band-structure calculations. $\gamma_n$ is the normal-state specific heat coefficient, $\lambda_\mathrm{ep}$ represents  the electron-phonon coupling constant, and $N(E_\mathrm{F})$ is the density of states at $E_\mathrm{F}$.
		\label{tab:parameter}}
	\begin{ruledtabular}
		\begin{tabular}{lccc}
			Property                                 & Unit               & NbRuSi        & TaRuSi    \\ \hline
			$T_c^\rho$                               & K                  & 3.0           & 4.1            \\
			$T_c^\chi$                               & K                  & 3.1           & 4.0            \\
			$T_c^C$                                  & K                  & 2.95          & 3.9            \\
			$T_c^{\mu\mathrm{SR}}$                   & K                  & 3.03          & 3.95            \\
			$\mu_0H_{c1}$                            & mT                 & 3.1(1)        & 5.6(1)         \\
			$\mu_0H_{c2}$                            & T                  & 1.40(5)       & 3.20(5)        \\
			$\gamma_n$                               & mJ/mol K$^{2}$       & 9.4(1)        & 8.4(2)        \\
			$N(E_\mathrm{F})^C$               & states/eV f.u.     & 3.98(4)       & 3.56(8)         \\
			$N(E_\mathrm{F})^\mathrm{DFT}$    & states/eV f.u.     & 2.67            & 1.81            \\
			$\lambda_\mathrm{ep}$                    & --                 & 0.46(1)           & 0.56(1)            \\
			$\lambda_\mathrm{12}$(=$\lambda_\mathrm{21}$)                    & --                 & 0.043          & 0.06            \\
			$\lambda_\mathrm{11}$(=$\lambda_\mathrm{22}$)                    & --                 & 0.22           & 0.29           \\
			$\lambda_0$\footnotemark[1]                               & nm                 & 303(2)        & 244(2)            \\
			$\Delta_0$({\textmu}SR)\footnotemark[1]                   & meV                & 0.47(1)        & 0.63(2)           \\
			$\Delta_{0,1}$({\textmu}SR)\footnotemark[2]                & meV                & 0.44(2)       & 0.58(3)           \\  
			$\Delta_{0,2}$({\textmu}SR)\footnotemark[2]                & meV                & 0.55(2)       & 0.68(3)           \\ 
			$w$ \footnotemark[3]                                              & --                  & 0.7          & 0.7            \\ 		    
			$\lambda_0$\footnotemark[3]                     & nm                 & 338(2)        & 216(2)            \\
			$\xi_1$\footnotemark[3]                         & nm                 & 19(1)         & 18(1)             \\
			$\xi_2$\footnotemark[3]                         & nm                 & 13(1)         & 10.5(5)             \\			 
			$\lambda_\mathrm{GL}(0)$                 & nm                 & 454(9)         & 344(3)            \\
			$\xi(0)$                                 & nm                 & 15.3(3)        & 10.1(1)           \\
			$\kappa$\footnotemark[4]                 & --                 & 30(1)          & 34(1)             \\
		\end{tabular}	
		\footnotetext[1]{Determined by the analysis of $\lambda_\mathrm{eff}^{-2}(T)$ using a one-gap $s$-wave model.}
		\footnotetext[2]{Determined by the analysis of $\lambda_\mathrm{eff}^{-2}(T)$ using a two-gap $s$-wave model.}
		\footnotetext[3]{Determined by the analysis of $\sigma_\mathrm{sc}(H)$ at 0.3\,K using a two-band model.}
		\footnotetext[4]{The Ginzburg-Landau parameter $\kappa$ was calculated  $\kappa$ = $\lambda_\mathrm{GL}(0)$/$\xi(0)$.}
	\end{ruledtabular}
\end{table}
%=== end table ==========================%

%==== figure =============================%
\begin{figure*}[!htp]
	\centering
	\includegraphics[width=0.8\textwidth,angle=0]{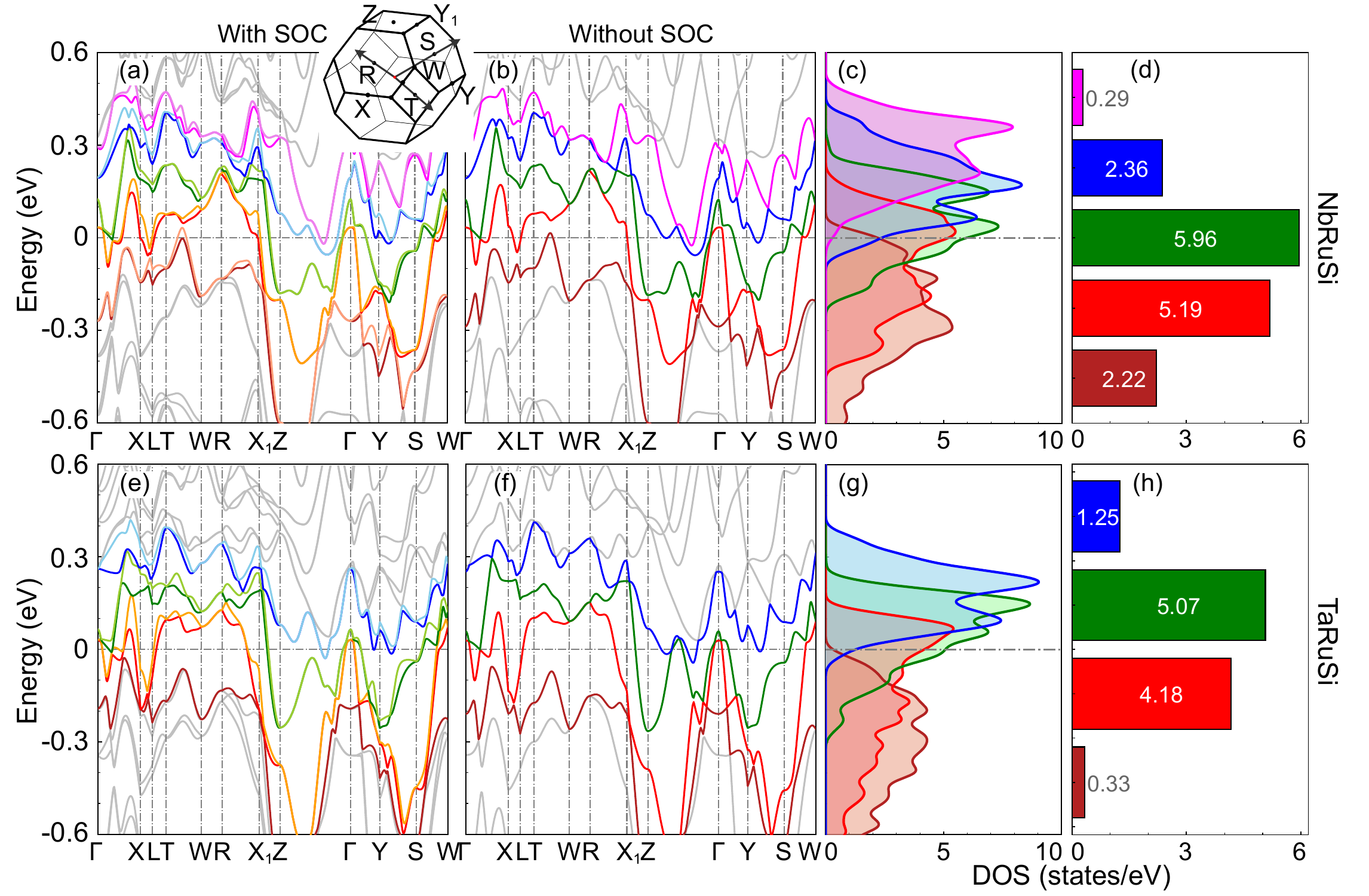}
	% \vspace{-2ex}%
	\caption{\label{fig:band}%
		Electronic band structures of NbRuSi calculated by considering (a) and by ignoring (b) the spin-orbit coupling. The bands that cross the Fermi level are highlighted using different colors. (c) Density of states contributed by each band shown in panel (b) for NbRuSi. (d)  Density of states contributed by each band at the Fermi level for NbRuSi. The analog results for TaRuSi are shown in panels (e)-(h). The bands that cross the Fermi level are highlighted using different colors.}
\end{figure*}
%=== end figure ==========================%	
%

By using the same weight $w$ as in $\sigma_\mathrm{sc}(H)$, 
we analyzed the tem\-pe\-ra\-ture-dependent superfluid
density $\rho_\mathrm{sc}(T)$ [$\propto \lambda_\mathrm{eff}^{-2}(T)$], 
now with a two-gap model. Here, we simply applied the $s$-wave model (i.e., singlet pairing) to analyze $\lambda_\mathrm{eff}^{-2}(T)$.
For simplicity, here we ignore the influence of the Fermi-surface
shape on the magnetic penetration depth and assume an isotropic
spherical Fermi surface for the fully-gapped superconductor.
Note that the ($s + ip$) model (i.e., a mixture of singlet- and triplet pairing) would have lead to similar results~\cite{Shang2022}.
As shown by lines in Fig.~\ref{fig:lam}, both one-gap- and two-gap models fit the experimental data comparably well. For the two-gap $s$-wave model,
the superfluid  density can be written as  $\rho_\mathrm{sc}(T) = w \rho_\mathrm{sc}^{\Delta_{0,1}}(T) + (1-w) \rho_\mathrm{sc}^{\Delta_{0,2}}(T)$, with $\rho_\mathrm{sc}^{\Delta_{0,1}}$ and $\rho_\mathrm{sc}^{\Delta_{0,2}}(T)$ 
the superfluid densities related to the first- ($\Delta_{0,1}$) and second ($\Delta_{0,2}$) gap, and $w$ a relative weight. 
Here, by fixing the weight $w$ to 0.7, the two-gap model provides almost identical results to the single-gap $s$-wave model, reflected in the two practically overlapping fitting curves shown in Fig.~\ref{fig:lam}. The two-gap model yields $\Delta_{0,1}$ = 0.44(2 )\,meV and $\Delta_{0,2}$ = 0.55(2)\,meV for NbRuSi, and $\Delta_{0,1}$ = 0.58(3)\,meV and $\Delta_{0,2}$ = 0.68(3)\,meV for TaRuSi, respectively.
For both compounds, the gap sizes are not significantly different
($\Delta_{0,1}$/$\Delta_{0,2}$ $\sim$ 0.80-0.85). This illustrates
clearly why it is so difficult to discriminate
between one-gap- and two-gap superconductors based on the temperature-dependent superfluid density alone~\cite{Khasanov2014,Khasanov2020,Shang2021}.  
At the same time, as we show above, the two-gap or multiband features
are clearly reflected in the field-dependent superconducting relaxation rate 
$\sigma_\mathrm{sc}(H)$ and also in the temperature-dependent upper critical field $H_\mathrm{c2}(T)$.

%==== figure =============================%
\begin{figure}[b]
	\centering
	\includegraphics[width=0.4\textwidth,angle=0]{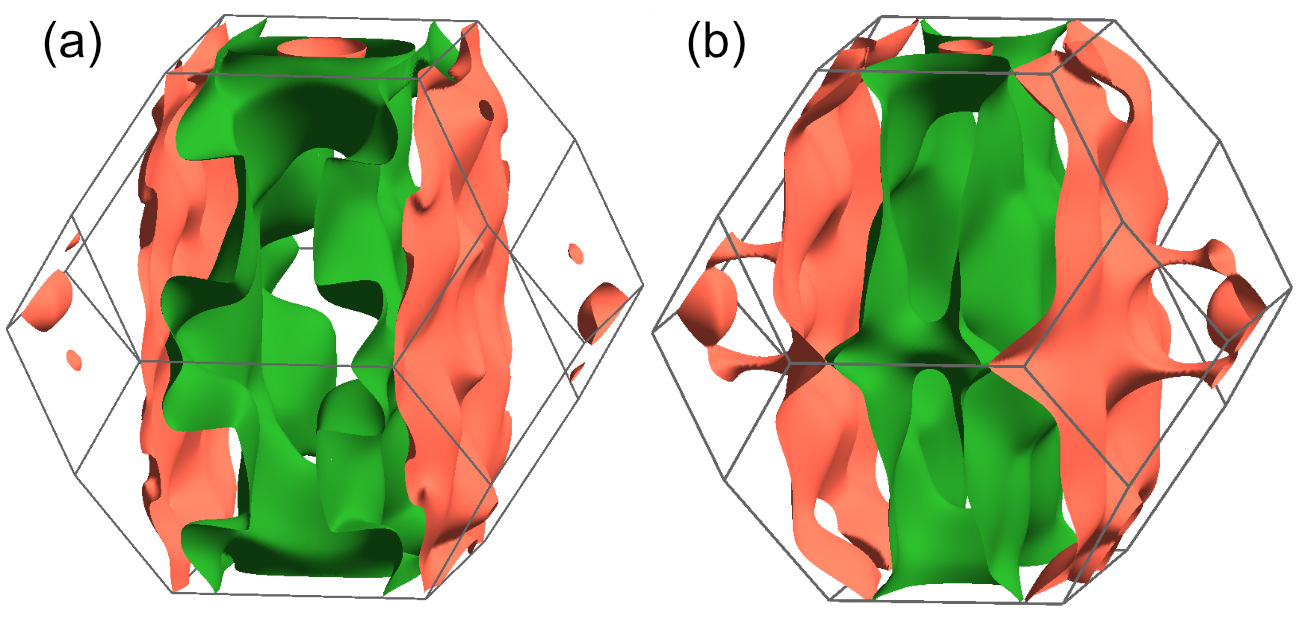}
	% \vspace{-2ex}%
	\caption{\label{fig:FS}%
	Representative Fermi surfaces for NbRuSi (a) and TaRuSi (b).
		In both cases, only the surfaces arising from the two
		dominant electronic bands (red and
		green bands in Fig.~\ref{fig:band}) are shown.}
\end{figure}
%=== end figure ==========================%	
%%%%%%%%%%%%%%%%%%%%%%%%%%%%%%%%%%%%%%%%%%%%

According to the temperature-dependent TF-{\textmu}SR measurements, the 
superfluid density $\rho_\mathrm{sc}(T)$ shows an almost tem\-per\-a\-ture\--independent behavior below 1/3$T_c$, indicating the absence of low-energy 
excitations and thus, a nodeless SC in both NbRuSi and TaRuSi.  
According to our previous work~\cite{Shang2022}, $\rho_\mathrm{sc}(T)$ is well described by an asymmetric ($s + ip$) pairing in both the weak- and the strong SOC limit across the full temperature range. In the weak SOC limit, the ($s + ip$)
pairing is identical to the conventional $s$-wave pairing. While in the strong SOC limit, the ($s + ip$) pairing implies a mixture of spin singlets and spin triplets, which is more consistent with a broken TRS superconducting state in both NbRuSi and TaRuSi. Note that, the above two limits lead to comparable results from the analysis of the $\rho_\mathrm{sc}(T)$ data. Both one-gap and two-gap $s$-wave models describe the $\rho_\mathrm{sc}(T)$ data 
very well (see Fig.~\ref{fig:lam}). However, the field-dependent superconducting Gaussian 
relaxation rate $\sigma_\mathrm{sc}(H)$ presented in this work (see 
Fig.~\ref{fig:relaxation}) provides clear evidence of multigap SC in both NbRuSi and TaRuSi, which shows a distinct field response compared to a single-gap 
superconductor~\cite{Shang2019c,Shang2020MoPB,Shang2021}. As shown in Fig.~\ref{fig:relaxation}, the single-band model deviates significantly from the measured $\sigma_\mathrm{sc}(H)$ data. By contrast, the two-band model yields upper critical fields 
$\mu_0H_{c2}$(0.3\,K) = 1.9(3) and 3.0(3)\,T for NbRuSi and TaRuSi, respectively, both comparable with the value determined from other techniques (see Fig.~\ref{fig:Hc2}). 
The derived \emph{virtual} upper critical fields  $\mu_0H_{c2}^\ast$ = 0.9(1) and 0.8(1)\,T for NbRuSi and TaRuSi 
correspond to the critical field value which suppresses the small superconducting energy gap. Clearly, the upper critical fields of both NbRuSi and TaRuSi are lower than the Pauli limiting field, suggesting that the orbital pair breaking is
the dominant mechanism in both compounds. 

The multigap SC of NbRuSi and TaRuSi can be further 
inferred from the temperature-dependent upper critical field $H_\mathrm{c2}(T)$. 
As shown in Fig.~\ref{fig:Hc2}, the two-band model is clearly superior to 
the WHH model over the whole temperature range, the latter being known to describe
very well the $H_\mathrm{c2}(T)$ of single-gap superconductors. 
The two-band model gives the intra-band and inter-band couplings $\lambda_{11}$ $\sim$ $\lambda_{22} = 0.22$ and $\lambda_{12} = 0.043$ for NbRuSi, 
and $\lambda_{11}$ $\sim$ $\lambda_{22} = 0.29$ and $\lambda_{12} = 0.06$ for TaRuSi, respectively. For both compounds, the intra-band coupling is almost 5 times larger than the inter-band coupling, which is different from the case of ZrNiAl-type NbReSi superconductor~\cite{Shang2022b}. In the latter case, the inter-band coupling is almost comparable to the intra-band coupling.   
As a consequence, the positive curvature in the $H_\mathrm{c2}(T)$ data, an indication of multiband SC, is more evident in NbRuSi and TaRuSi than in NbReSi. Similarly to NbReSi, the multiband features are also less evident in the TiFeSi-type TaReSi, isostructural to NbRuSi and TaRuSi~\cite{Shang2023}. Therefore, we expect comparable inter- and intra-band couplings also in the TaReSi superconductor. 

Finally, the multiband SC is further supported by electronic band-structure calculations. 
As shown in Fig.~\ref{fig:band}(b) and (f), up to five (four) bands are identified to cross the Fermi level of NbRuSi (TaRuSi), confirming their multiband nature. 
When taking into account the SOC effects [see Fig.~\ref{fig:band}(a) and (e)], there are even more bands that cross the Fermi level: ten and seven in NbRuSi and TaRuSi, respectively. The calculated density of states (DOS) (without considering SOC) of both compounds is comparable with the value determined from normal-state electronic specific-heat coefficient [see Table~\ref{tab:parameter} and Fig.~\ref{fig:band}(c) and (g)]. 
The density of states at the Fermi level arises mostly of contributions from the Nb-$4d$ (Ta-$5d$), Ru-$4d$ and Si-$3p$ orbitals.
Due to the larger atomic number of Ta, the band splitting in TaRuSi is clearly larger than in NbRuSi. For instance, at the high-symmetry $X$ point, the band splitting is about 128 and 79\,meV in TaRuSi and NbRuSi, respectively. We also estimated the contributions of each band to the DOS at the Fermi level. In NbRuSi, out of a total of five bands,  two of them contribute almost 70\% of the DOS [see green and red bars in Fig.~\ref{fig:band}(d)]. In the TaRuSi case, two bands (out of four bands) contribute over 85\% of the total DOS [see Fig.~\ref{fig:band}(h)]. Interestingly, for both compounds, the two
dominant electronic bands exhibit quasi-two-dimensional Fermi surfaces (see Fig.~\ref{fig:FS}).
Clearly, since at least two bands contribute almost equally to the DOS at Fermi level (and, hence, to the electronic properties), this is highly consistent with the observation of multiband SC in both NbRuSi and TaRuSi superconductors.  

\section{\label{ssec:Sum}Conclusion}\enlargethispage{8pt}
To summarize, we investigated the multiband nature of the NbRuSi
and TaRuSi unconventional superconductors by means of TF-{\textmu}SR,
electrical-resistivity and heat-capacity measurements. NbRuSi and TaRuSi undergo a bulk superconducting transition at $T_c =  $ 3 and 4\,K, respectively. The previous zero-field {\textmu}SR measurements reveal that both compounds spontaneously break the time-reversal symmetry at the superconducting transition, while unexpectedly showing a fully gapped SC.
In both cases, the temperature-dependent superfluid density obtained via TF-{\textmu}SR fails to distinguish a single-gap from a multigap scenario. At the same time, their multigap features are clearly reflected in the field-dependent superconducting Gaussian relaxation rate $\sigma_\mathrm{sc}(H)$ and the temperature-dependent upper critical field $H_\mathrm{c2}(T)$. By combining the experimental results presented in this work with numerical band-structure calculations, we provide solid evidence for the multigap SC in both NbRuSi and TaRuSi, and thus, offer new insight into their unconventional or topological superconducting nature.\\

\vspace{1pt}
\begin{acknowledgments}
This work was supported by the National Natural Science Foundation of China (Grant Nos. 12374105 and 12350710785), the 
Natural Science Foundation of Shanghai (Grant Nos.\ 21ZR1420500 and 21JC\-140\-2300), the Natural Science Foundation
of Chongqing (Grant No.\ CSTB-2022NSCQ-MSX1678), and the Fundamental Research Funds for the Central Universities.  
We acknowledge the allocation of beam time at the Swiss muon source
(Dolly {\textmu}SR spectrometer).
\end{acknowledgments}

%\appendix
%\section{\label{appendix} Frequency detuning of the NMR resonant circuit}

%\begin{footnotesize}
\bibliography{TaRuSi.bib}

\begin{thebibliography}{70}%
\makeatletter
\providecommand \@ifxundefined [1]{%
 \@ifx{#1\undefined}
}%
\providecommand \@ifnum [1]{%
 \ifnum #1\expandafter \@firstoftwo
 \else \expandafter \@secondoftwo
 \fi
}%
\providecommand \@ifx [1]{%
 \ifx #1\expandafter \@firstoftwo
 \else \expandafter \@secondoftwo
 \fi
}%
\providecommand \natexlab [1]{#1}%
\providecommand \enquote  [1]{``#1''}%
\providecommand \bibnamefont  [1]{#1}%
\providecommand \bibfnamefont [1]{#1}%
\providecommand \citenamefont [1]{#1}%
\providecommand \href@noop [0]{\@secondoftwo}%
\providecommand \href [0]{\begingroup \@sanitize@url \@href}%
\providecommand \@href[1]{\@@startlink{#1}\@@href}%
\providecommand \@@href[1]{\endgroup#1\@@endlink}%
\providecommand \@sanitize@url [0]{\catcode `\\12\catcode `\$12\catcode
  `\&12\catcode `\#12\catcode `\^12\catcode `\_12\catcode `\%12\relax}%
\providecommand \@@startlink[1]{}%
\providecommand \@@endlink[0]{}%
\providecommand \url  [0]{\begingroup\@sanitize@url \@url }%
\providecommand \@url [1]{\endgroup\@href {#1}{\urlprefix }}%
\providecommand \urlprefix  [0]{URL }%
\providecommand \Eprint [0]{\href }%
\providecommand \doibase [0]{https://doi.org/}%
\providecommand \selectlanguage [0]{\@gobble}%
\providecommand \bibinfo  [0]{\@secondoftwo}%
\providecommand \bibfield  [0]{\@secondoftwo}%
\providecommand \translation [1]{[#1]}%
\providecommand \BibitemOpen [0]{}%
\providecommand \bibitemStop [0]{}%
\providecommand \bibitemNoStop [0]{.\EOS\space}%
\providecommand \EOS [0]{\spacefactor3000\relax}%
\providecommand \BibitemShut  [1]{\csname bibitem#1\endcsname}%
\let\auto@bib@innerbib\@empty
%</preamble>
\bibitem [{\citenamefont {Armitage}\ \emph {et~al.}(2018)\citenamefont
  {Armitage}, \citenamefont {Mele},\ and\ \citenamefont
  {Vishwanath}}]{Armitage2018}%
  \BibitemOpen
  \bibfield  {author} {\bibinfo {author} {\bibfnamefont {N.~P.}\ \bibnamefont
  {Armitage}}, \bibinfo {author} {\bibfnamefont {E.~J.}\ \bibnamefont {Mele}},\
  and\ \bibinfo {author} {\bibfnamefont {A.}~\bibnamefont {Vishwanath}},\
  }\bibfield  {title} {\bibinfo {title} {Weyl and {Dirac} semimetals in
  three-dimensional solids},\ }\href
  {https://doi.org/10.1103/RevModPhys.90.015001} {\bibfield  {journal}
  {\bibinfo  {journal} {Rev. Mod. Phys.}\ }\textbf {\bibinfo {volume} {90}},\
  \bibinfo {pages} {015001} (\bibinfo {year} {2018})}\BibitemShut {NoStop}%
\bibitem [{\citenamefont {Lv}\ \emph {et~al.}(2021)\citenamefont {Lv},
  \citenamefont {Qian},\ and\ \citenamefont {Ding}}]{Lv2021}%
  \BibitemOpen
  \bibfield  {author} {\bibinfo {author} {\bibfnamefont {B.~Q.}\ \bibnamefont
  {Lv}}, \bibinfo {author} {\bibfnamefont {T.}~\bibnamefont {Qian}},\ and\
  \bibinfo {author} {\bibfnamefont {H.}~\bibnamefont {Ding}},\ }\bibfield
  {title} {\bibinfo {title} {Experimental perspective on three-dimensional
  topological semimetals},\ }\href
  {https://doi.org/10.1103/RevModPhys.93.025002} {\bibfield  {journal}
  {\bibinfo  {journal} {Rev. Mod. Phys.}\ }\textbf {\bibinfo {volume} {93}},\
  \bibinfo {pages} {025002} (\bibinfo {year} {2021})}\BibitemShut {NoStop}%
\bibitem [{\citenamefont {Yan}\ and\ \citenamefont {Felser}(2017)}]{Yan2017}%
  \BibitemOpen
  \bibfield  {author} {\bibinfo {author} {\bibfnamefont {B.}~\bibnamefont
  {Yan}}\ and\ \bibinfo {author} {\bibfnamefont {C.}~\bibnamefont {Felser}},\
  }\bibfield  {title} {\bibinfo {title} {Topological materials: {Weyl}
  semimetals},\ }\href
  {https://doi.org/10.1146/annurev-conmatphys-031016-025458} {\bibfield
  {journal} {\bibinfo  {journal} {Annu. Rev. Condens. Matter Phys.}\ }\textbf
  {\bibinfo {volume} {8}},\ \bibinfo {pages} {337} (\bibinfo {year}
  {2017})}\BibitemShut {NoStop}%
\bibitem [{\citenamefont {Wieder}\ \emph {et~al.}(2022)\citenamefont {Wieder},
  \citenamefont {Bradlyn}, \citenamefont {Cano}, \citenamefont {Wang},
  \citenamefont {Vergniory}, \citenamefont {Elcoro}, \citenamefont {Soluyanov},
  \citenamefont {Felser}, \citenamefont {Neupert}, \citenamefont {Regnault},\
  and\ \citenamefont {Bernevig}}]{Wieder2022}%
  \BibitemOpen
  \bibfield  {author} {\bibinfo {author} {\bibfnamefont {B.~J.}\ \bibnamefont
  {Wieder}}, \bibinfo {author} {\bibfnamefont {B.}~\bibnamefont {Bradlyn}},
  \bibinfo {author} {\bibfnamefont {J.}~\bibnamefont {Cano}}, \bibinfo {author}
  {\bibfnamefont {Z.}~\bibnamefont {Wang}}, \bibinfo {author} {\bibfnamefont
  {M.~G.}\ \bibnamefont {Vergniory}}, \bibinfo {author} {\bibfnamefont
  {L.}~\bibnamefont {Elcoro}}, \bibinfo {author} {\bibfnamefont {A.~A.}\
  \bibnamefont {Soluyanov}}, \bibinfo {author} {\bibfnamefont {C.}~\bibnamefont
  {Felser}}, \bibinfo {author} {\bibfnamefont {T.}~\bibnamefont {Neupert}},
  \bibinfo {author} {\bibfnamefont {N.}~\bibnamefont {Regnault}},\ and\
  \bibinfo {author} {\bibfnamefont {B.~A.}\ \bibnamefont {Bernevig}},\
  }\bibfield  {title} {\bibinfo {title} {Topological materials discovery from
  crystal symmetry},\ }\href {https://doi.org/10.1038/s41578-021-00380-2}
  {\bibfield  {journal} {\bibinfo  {journal} {Nat. Rev. Mater.}\ }\textbf
  {\bibinfo {volume} {7}},\ \bibinfo {pages} {196} (\bibinfo {year}
  {2022})}\BibitemShut {NoStop}%
\bibitem [{\citenamefont {Bradlyn}\ \emph {et~al.}(2017)\citenamefont
  {Bradlyn}, \citenamefont {Elcoro}, \citenamefont {Cano}, \citenamefont
  {Vergniory}, \citenamefont {Wang}, \citenamefont {Felser}, \citenamefont
  {Aroyo},\ and\ \citenamefont {Bernevig}}]{Bradlyn2017}%
  \BibitemOpen
  \bibfield  {author} {\bibinfo {author} {\bibfnamefont {B.}~\bibnamefont
  {Bradlyn}}, \bibinfo {author} {\bibfnamefont {L.}~\bibnamefont {Elcoro}},
  \bibinfo {author} {\bibfnamefont {J.}~\bibnamefont {Cano}}, \bibinfo {author}
  {\bibfnamefont {M.~G.}\ \bibnamefont {Vergniory}}, \bibinfo {author}
  {\bibfnamefont {Z.}~\bibnamefont {Wang}}, \bibinfo {author} {\bibfnamefont
  {C.}~\bibnamefont {Felser}}, \bibinfo {author} {\bibfnamefont {M.~I.}\
  \bibnamefont {Aroyo}},\ and\ \bibinfo {author} {\bibfnamefont {B.~A.}\
  \bibnamefont {Bernevig}},\ }\bibfield  {title} {\bibinfo {title} {Topological
  quantum chemistry},\ }\href {https://doi.org/10.1038/nature23268} {\bibfield
  {journal} {\bibinfo  {journal} {Nature}\ }\textbf {\bibinfo {volume} {547}},\
  \bibinfo {pages} {298} (\bibinfo {year} {2017})}\BibitemShut {NoStop}%
\bibitem [{\citenamefont {Vergniory}\ \emph {et~al.}(2019)\citenamefont
  {Vergniory}, \citenamefont {Elcoro}, \citenamefont {Felser}, \citenamefont
  {Regnault}, \citenamefont {Bernevig},\ and\ \citenamefont
  {Wang}}]{Vergniory2019}%
  \BibitemOpen
  \bibfield  {author} {\bibinfo {author} {\bibfnamefont {M.~G.}\ \bibnamefont
  {Vergniory}}, \bibinfo {author} {\bibfnamefont {L.}~\bibnamefont {Elcoro}},
  \bibinfo {author} {\bibfnamefont {C.}~\bibnamefont {Felser}}, \bibinfo
  {author} {\bibfnamefont {N.}~\bibnamefont {Regnault}}, \bibinfo {author}
  {\bibfnamefont {B.~A.}\ \bibnamefont {Bernevig}},\ and\ \bibinfo {author}
  {\bibfnamefont {Z.}~\bibnamefont {Wang}},\ }\bibfield  {title} {\bibinfo
  {title} {A complete catalogue of high-quality topological materials},\ }\href
  {https://doi.org/10.1038/s41586-019-0954-4} {\bibfield  {journal} {\bibinfo
  {journal} {Nature}\ }\textbf {\bibinfo {volume} {566}},\ \bibinfo {pages}
  {480} (\bibinfo {year} {2019})}\BibitemShut {NoStop}%
\bibitem [{\citenamefont {Tang}\ \emph {et~al.}(2019)\citenamefont {Tang},
  \citenamefont {Po}, \citenamefont {Vishwanath},\ and\ \citenamefont
  {Wan}}]{Tang2019}%
  \BibitemOpen
  \bibfield  {author} {\bibinfo {author} {\bibfnamefont {F.}~\bibnamefont
  {Tang}}, \bibinfo {author} {\bibfnamefont {H.~C.}\ \bibnamefont {Po}},
  \bibinfo {author} {\bibfnamefont {A.}~\bibnamefont {Vishwanath}},\ and\
  \bibinfo {author} {\bibfnamefont {X.}~\bibnamefont {Wan}},\ }\bibfield
  {title} {\bibinfo {title} {Comprehensive search for topological materials
  using symmetry indicators},\ }\href
  {https://doi.org/10.1038/s41586-019-0937-5} {\bibfield  {journal} {\bibinfo
  {journal} {Nature}\ }\textbf {\bibinfo {volume} {566}},\ \bibinfo {pages}
  {486} (\bibinfo {year} {2019})}\BibitemShut {NoStop}%
\bibitem [{\citenamefont {Bernevig}\ \emph {et~al.}(2022)\citenamefont
  {Bernevig}, \citenamefont {Felser},\ and\ \citenamefont
  {Beidenkopf}}]{Bernevig2022}%
  \BibitemOpen
  \bibfield  {author} {\bibinfo {author} {\bibfnamefont {B.~A.}\ \bibnamefont
  {Bernevig}}, \bibinfo {author} {\bibfnamefont {C.}~\bibnamefont {Felser}},\
  and\ \bibinfo {author} {\bibfnamefont {H.}~\bibnamefont {Beidenkopf}},\
  }\bibfield  {title} {\bibinfo {title} {Progress and prospects in magnetic
  topological materials},\ }\href {https://doi.org/10.1038/s41586-021-04105-x}
  {\bibfield  {journal} {\bibinfo  {journal} {Nature}\ }\textbf {\bibinfo
  {volume} {603}},\ \bibinfo {pages} {41} (\bibinfo {year} {2022})}\BibitemShut
  {NoStop}%
\bibitem [{\citenamefont {Xu}\ \emph {et~al.}(2015{\natexlab{a}})\citenamefont
  {Xu}, \citenamefont {Belopolski}, \citenamefont {Alidoust}, \citenamefont
  {Neupane}, \citenamefont {Bian}, \citenamefont {Zhang}, \citenamefont
  {Sankar}, \citenamefont {Chang}, \citenamefont {Yuan}, \citenamefont {Lee},
  \citenamefont {Huang}, \citenamefont {Zheng}, \citenamefont {Ma},
  \citenamefont {Sanchez}, \citenamefont {Wang}, \citenamefont {Bansil},
  \citenamefont {Chou}, \citenamefont {Shibayev}, \citenamefont {Lin},
  \citenamefont {Jia},\ and\ \citenamefont {Hasan}}]{Xu2015a}%
  \BibitemOpen
  \bibfield  {author} {\bibinfo {author} {\bibfnamefont {S.-Y.}\ \bibnamefont
  {Xu}}, \bibinfo {author} {\bibfnamefont {I.}~\bibnamefont {Belopolski}},
  \bibinfo {author} {\bibfnamefont {N.}~\bibnamefont {Alidoust}}, \bibinfo
  {author} {\bibfnamefont {M.}~\bibnamefont {Neupane}}, \bibinfo {author}
  {\bibfnamefont {G.}~\bibnamefont {Bian}}, \bibinfo {author} {\bibfnamefont
  {C.}~\bibnamefont {Zhang}}, \bibinfo {author} {\bibfnamefont
  {R.}~\bibnamefont {Sankar}}, \bibinfo {author} {\bibfnamefont
  {G.}~\bibnamefont {Chang}}, \bibinfo {author} {\bibfnamefont
  {Z.}~\bibnamefont {Yuan}}, \bibinfo {author} {\bibfnamefont {C.-C.}\
  \bibnamefont {Lee}}, \bibinfo {author} {\bibfnamefont {S.-M.}\ \bibnamefont
  {Huang}}, \bibinfo {author} {\bibfnamefont {H.}~\bibnamefont {Zheng}},
  \bibinfo {author} {\bibfnamefont {J.}~\bibnamefont {Ma}}, \bibinfo {author}
  {\bibfnamefont {D.~S.}\ \bibnamefont {Sanchez}}, \bibinfo {author}
  {\bibfnamefont {B.}~\bibnamefont {Wang}}, \bibinfo {author} {\bibfnamefont
  {A.}~\bibnamefont {Bansil}}, \bibinfo {author} {\bibfnamefont
  {F.}~\bibnamefont {Chou}}, \bibinfo {author} {\bibfnamefont {P.~P.}\
  \bibnamefont {Shibayev}}, \bibinfo {author} {\bibfnamefont {H.}~\bibnamefont
  {Lin}}, \bibinfo {author} {\bibfnamefont {S.}~\bibnamefont {Jia}},\ and\
  \bibinfo {author} {\bibfnamefont {M.~Z.}\ \bibnamefont {Hasan}},\ }\bibfield
  {title} {\bibinfo {title} {Discovery of a {Weyl} fermion semimetal and
  topological {Fermi} arcs},\ }\href {https://doi.org/10.1126/science.aaa9297}
  {\bibfield  {journal} {\bibinfo  {journal} {Science}\ }\textbf {\bibinfo
  {volume} {349}},\ \bibinfo {pages} {613} (\bibinfo {year}
  {2015}{\natexlab{a}})}\BibitemShut {NoStop}%
\bibitem [{\citenamefont {Xu}\ \emph {et~al.}(2015{\natexlab{b}})\citenamefont
  {Xu}, \citenamefont {Alidoust}, \citenamefont {Belopolski}, \citenamefont
  {Yuan}, \citenamefont {Bian}, \citenamefont {Chang}, \citenamefont {Zheng},
  \citenamefont {Strocov}, \citenamefont {Sanchez}, \citenamefont {Chang} \emph
  {et~al.}}]{Xu2015b}%
  \BibitemOpen
  \bibfield  {author} {\bibinfo {author} {\bibfnamefont {S.-Y.}\ \bibnamefont
  {Xu}}, \bibinfo {author} {\bibfnamefont {N.}~\bibnamefont {Alidoust}},
  \bibinfo {author} {\bibfnamefont {I.}~\bibnamefont {Belopolski}}, \bibinfo
  {author} {\bibfnamefont {Z.}~\bibnamefont {Yuan}}, \bibinfo {author}
  {\bibfnamefont {G.}~\bibnamefont {Bian}}, \bibinfo {author} {\bibfnamefont
  {T.-R.}\ \bibnamefont {Chang}}, \bibinfo {author} {\bibfnamefont
  {H.}~\bibnamefont {Zheng}}, \bibinfo {author} {\bibfnamefont {V.~N.}\
  \bibnamefont {Strocov}}, \bibinfo {author} {\bibfnamefont {D.~S.}\
  \bibnamefont {Sanchez}}, \bibinfo {author} {\bibfnamefont {G.}~\bibnamefont
  {Chang}}, \emph {et~al.},\ }\bibfield  {title} {\bibinfo {title} {Discovery
  of a {Weyl} fermion state with {Fermi} arcs in niobium arsenide},\ }\href
  {https://doi.org/10.1038/nphys3437} {\bibfield  {journal} {\bibinfo
  {journal} {Nat. Phys.}\ }\textbf {\bibinfo {volume} {11}},\ \bibinfo {pages}
  {748} (\bibinfo {year} {2015}{\natexlab{b}})}\BibitemShut {NoStop}%
\bibitem [{\citenamefont {Lv}\ \emph {et~al.}(2015)\citenamefont {Lv},
  \citenamefont {Weng}, \citenamefont {Fu}, \citenamefont {Wang}, \citenamefont
  {Miao}, \citenamefont {Ma}, \citenamefont {Richard}, \citenamefont {Huang},
  \citenamefont {Zhao}, \citenamefont {Chen}, \citenamefont {Fang},
  \citenamefont {Dai}, \citenamefont {Qian},\ and\ \citenamefont
  {Ding}}]{Lv2015}%
  \BibitemOpen
  \bibfield  {author} {\bibinfo {author} {\bibfnamefont {B.~Q.}\ \bibnamefont
  {Lv}}, \bibinfo {author} {\bibfnamefont {H.~M.}\ \bibnamefont {Weng}},
  \bibinfo {author} {\bibfnamefont {B.~B.}\ \bibnamefont {Fu}}, \bibinfo
  {author} {\bibfnamefont {X.~P.}\ \bibnamefont {Wang}}, \bibinfo {author}
  {\bibfnamefont {H.}~\bibnamefont {Miao}}, \bibinfo {author} {\bibfnamefont
  {J.}~\bibnamefont {Ma}}, \bibinfo {author} {\bibfnamefont {P.}~\bibnamefont
  {Richard}}, \bibinfo {author} {\bibfnamefont {X.~C.}\ \bibnamefont {Huang}},
  \bibinfo {author} {\bibfnamefont {L.~X.}\ \bibnamefont {Zhao}}, \bibinfo
  {author} {\bibfnamefont {G.~F.}\ \bibnamefont {Chen}}, \bibinfo {author}
  {\bibfnamefont {Z.}~\bibnamefont {Fang}}, \bibinfo {author} {\bibfnamefont
  {X.}~\bibnamefont {Dai}}, \bibinfo {author} {\bibfnamefont {T.}~\bibnamefont
  {Qian}},\ and\ \bibinfo {author} {\bibfnamefont {H.}~\bibnamefont {Ding}},\
  }\bibfield  {title} {\bibinfo {title} {Experimental discovery of {W}eyl
  semimetal {TaAs}},\ }\href {https://doi.org/10.1103/PhysRevX.5.031013}
  {\bibfield  {journal} {\bibinfo  {journal} {Phys. Rev. X}\ }\textbf {\bibinfo
  {volume} {5}},\ \bibinfo {pages} {031013} (\bibinfo {year}
  {2015})}\BibitemShut {NoStop}%
\bibitem [{\citenamefont {Xu}\ \emph {et~al.}(2016)\citenamefont {Xu},
  \citenamefont {Weng}, \citenamefont {Lv}, \citenamefont {Matt}, \citenamefont
  {Park}, \citenamefont {Bisti}, \citenamefont {Strocov}, \citenamefont
  {Gawryluk}, \citenamefont {Pomjakushina}, \citenamefont {Conder},
  \citenamefont {Plumb}, \citenamefont {Radovic}, \citenamefont {Aut\'es},
  \citenamefont {Yazyev}, \citenamefont {Fang}, \citenamefont {Dai},
  \citenamefont {Qian}, \citenamefont {Mesot}, \citenamefont {Ding},\ and\
  \citenamefont {Shi}}]{Xu2016}%
  \BibitemOpen
  \bibfield  {author} {\bibinfo {author} {\bibfnamefont {N.}~\bibnamefont
  {Xu}}, \bibinfo {author} {\bibfnamefont {H.~M.}\ \bibnamefont {Weng}},
  \bibinfo {author} {\bibfnamefont {B.~Q.}\ \bibnamefont {Lv}}, \bibinfo
  {author} {\bibfnamefont {C.~E.}\ \bibnamefont {Matt}}, \bibinfo {author}
  {\bibfnamefont {J.}~\bibnamefont {Park}}, \bibinfo {author} {\bibfnamefont
  {F.}~\bibnamefont {Bisti}}, \bibinfo {author} {\bibfnamefont {V.~N.}\
  \bibnamefont {Strocov}}, \bibinfo {author} {\bibfnamefont {D.}~\bibnamefont
  {Gawryluk}}, \bibinfo {author} {\bibfnamefont {E.}~\bibnamefont
  {Pomjakushina}}, \bibinfo {author} {\bibfnamefont {K.}~\bibnamefont
  {Conder}}, \bibinfo {author} {\bibfnamefont {N.~C.}\ \bibnamefont {Plumb}},
  \bibinfo {author} {\bibfnamefont {M.}~\bibnamefont {Radovic}}, \bibinfo
  {author} {\bibfnamefont {G.}~\bibnamefont {Aut\'es}}, \bibinfo {author}
  {\bibfnamefont {O.~V.}\ \bibnamefont {Yazyev}}, \bibinfo {author}
  {\bibfnamefont {Z.}~\bibnamefont {Fang}}, \bibinfo {author} {\bibfnamefont
  {X.}~\bibnamefont {Dai}}, \bibinfo {author} {\bibfnamefont {T.}~\bibnamefont
  {Qian}}, \bibinfo {author} {\bibfnamefont {J.}~\bibnamefont {Mesot}},
  \bibinfo {author} {\bibfnamefont {H.}~\bibnamefont {Ding}},\ and\ \bibinfo
  {author} {\bibfnamefont {M.}~\bibnamefont {Shi}},\ }\bibfield  {title}
  {\bibinfo {title} {Observation of {Weyl} nodes and {Fermi} arcs in tantalum
  phosphide},\ }\href {https://doi.org/10.1038/ncomms11006} {\bibfield
  {journal} {\bibinfo  {journal} {Nat. Commun.}\ }\textbf {\bibinfo {volume}
  {7}},\ \bibinfo {pages} {11006} (\bibinfo {year} {2016})}\BibitemShut
  {NoStop}%
\bibitem [{\citenamefont {Souma}\ \emph {et~al.}(2016)\citenamefont {Souma},
  \citenamefont {Wang}, \citenamefont {Kotaka}, \citenamefont {Sato},
  \citenamefont {Nakayama}, \citenamefont {Tanaka}, \citenamefont {Kimizuka},
  \citenamefont {Takahashi}, \citenamefont {Yamauchi}, \citenamefont {Oguchi},
  \citenamefont {Segawa},\ and\ \citenamefont {Ando}}]{Souma2016}%
  \BibitemOpen
  \bibfield  {author} {\bibinfo {author} {\bibfnamefont {S.}~\bibnamefont
  {Souma}}, \bibinfo {author} {\bibfnamefont {Z.}~\bibnamefont {Wang}},
  \bibinfo {author} {\bibfnamefont {H.}~\bibnamefont {Kotaka}}, \bibinfo
  {author} {\bibfnamefont {T.}~\bibnamefont {Sato}}, \bibinfo {author}
  {\bibfnamefont {K.}~\bibnamefont {Nakayama}}, \bibinfo {author}
  {\bibfnamefont {Y.}~\bibnamefont {Tanaka}}, \bibinfo {author} {\bibfnamefont
  {H.}~\bibnamefont {Kimizuka}}, \bibinfo {author} {\bibfnamefont
  {T.}~\bibnamefont {Takahashi}}, \bibinfo {author} {\bibfnamefont
  {K.}~\bibnamefont {Yamauchi}}, \bibinfo {author} {\bibfnamefont
  {T.}~\bibnamefont {Oguchi}}, \bibinfo {author} {\bibfnamefont
  {K.}~\bibnamefont {Segawa}},\ and\ \bibinfo {author} {\bibfnamefont
  {Y.}~\bibnamefont {Ando}},\ }\bibfield  {title} {\bibinfo {title} {Direct
  observation of nonequivalent {Fermi}-arc states of opposite surfaces in the
  noncentrosymmetric {Weyl} semimetal {NbP}},\ }\href
  {https://doi.org/10.1103/PhysRevB.93.161112} {\bibfield  {journal} {\bibinfo
  {journal} {Phys. Rev. B}\ }\textbf {\bibinfo {volume} {93}},\ \bibinfo
  {pages} {161112(R)} (\bibinfo {year} {2016})}\BibitemShut {NoStop}%
\bibitem [{\citenamefont {Wang}\ \emph {et~al.}(2016)\citenamefont {Wang},
  \citenamefont {Alexandradinata}, \citenamefont {Cava},\ and\ \citenamefont
  {Bernevig}}]{Wang2016}%
  \BibitemOpen
  \bibfield  {author} {\bibinfo {author} {\bibfnamefont {Z.}~\bibnamefont
  {Wang}}, \bibinfo {author} {\bibfnamefont {A.}~\bibnamefont
  {Alexandradinata}}, \bibinfo {author} {\bibfnamefont {R.~J.}\ \bibnamefont
  {Cava}},\ and\ \bibinfo {author} {\bibfnamefont {B.~A.}\ \bibnamefont
  {Bernevig}},\ }\bibfield  {title} {\bibinfo {title} {Hourglass fermions},\
  }\href {https://doi.org/10.1038/nature17410} {\bibfield  {journal} {\bibinfo
  {journal} {Nature}\ }\textbf {\bibinfo {volume} {532}},\ \bibinfo {pages}
  {189} (\bibinfo {year} {2016})}\BibitemShut {NoStop}%
\bibitem [{\citenamefont {Wu}\ \emph {et~al.}(2019)\citenamefont {Wu},
  \citenamefont {Jiao}, \citenamefont {Li}, \citenamefont {Sheng},
  \citenamefont {Yu},\ and\ \citenamefont {Yang}}]{Wu2019}%
  \BibitemOpen
  \bibfield  {author} {\bibinfo {author} {\bibfnamefont {W.}~\bibnamefont
  {Wu}}, \bibinfo {author} {\bibfnamefont {Y.}~\bibnamefont {Jiao}}, \bibinfo
  {author} {\bibfnamefont {S.}~\bibnamefont {Li}}, \bibinfo {author}
  {\bibfnamefont {X.-L.}\ \bibnamefont {Sheng}}, \bibinfo {author}
  {\bibfnamefont {Z.-M.}\ \bibnamefont {Yu}},\ and\ \bibinfo {author}
  {\bibfnamefont {S.~A.}\ \bibnamefont {Yang}},\ }\bibfield  {title} {\bibinfo
  {title} {Hourglass {W}eyl loops in two dimensions: {T}heory and material
  realization in monolayer {GaTeI} family},\ }\href
  {https://doi.org/10.1103/physrevmaterials.3.054203} {\bibfield  {journal}
  {\bibinfo  {journal} {Phys. Rev. Mater.}\ }\textbf {\bibinfo {volume} {3}},\
  \bibinfo {pages} {054203} (\bibinfo {year} {2019})}\BibitemShut {NoStop}%
\bibitem [{\citenamefont {Shang}\ \emph
  {et~al.}(2022{\natexlab{a}})\citenamefont {Shang}, \citenamefont {Zhao},
  \citenamefont {Hu}, \citenamefont {Ma}, \citenamefont {Gawryluk},
  \citenamefont {Zhu}, \citenamefont {Zhang}, \citenamefont {Zhen},
  \citenamefont {Yu}, \citenamefont {Xu}, \citenamefont {Zhan}, \citenamefont
  {Pomjakushina}, \citenamefont {Shi},\ and\ \citenamefont
  {Shiroka}}]{Shang2022}%
  \BibitemOpen
  \bibfield  {author} {\bibinfo {author} {\bibfnamefont {T.}~\bibnamefont
  {Shang}}, \bibinfo {author} {\bibfnamefont {J.}~\bibnamefont {Zhao}},
  \bibinfo {author} {\bibfnamefont {L.-H.}\ \bibnamefont {Hu}}, \bibinfo
  {author} {\bibfnamefont {J.}~\bibnamefont {Ma}}, \bibinfo {author}
  {\bibfnamefont {D.~J.}\ \bibnamefont {Gawryluk}}, \bibinfo {author}
  {\bibfnamefont {X.}~\bibnamefont {Zhu}}, \bibinfo {author} {\bibfnamefont
  {H.}~\bibnamefont {Zhang}}, \bibinfo {author} {\bibfnamefont
  {Z.}~\bibnamefont {Zhen}}, \bibinfo {author} {\bibfnamefont {B.}~\bibnamefont
  {Yu}}, \bibinfo {author} {\bibfnamefont {Y.}~\bibnamefont {Xu}}, \bibinfo
  {author} {\bibfnamefont {Q.}~\bibnamefont {Zhan}}, \bibinfo {author}
  {\bibfnamefont {E.}~\bibnamefont {Pomjakushina}}, \bibinfo {author}
  {\bibfnamefont {M.}~\bibnamefont {Shi}},\ and\ \bibinfo {author}
  {\bibfnamefont {T.}~\bibnamefont {Shiroka}},\ }\bibfield  {title} {\bibinfo
  {title} {Unconventional superconductivity in topological {Kramers} nodal-line
  semimetals},\ }\href {https://doi.org/10.1126/sciadv.abq6589} {\bibfield
  {journal} {\bibinfo  {journal} {Sci. Adv.}\ }\textbf {\bibinfo {volume}
  {8}},\ \bibinfo {pages} {eabq6589} (\bibinfo {year}
  {2022}{\natexlab{a}})}\BibitemShut {NoStop}%
\bibitem [{\citenamefont {Shang}\ \emph {et~al.}(2023)\citenamefont {Shang},
  \citenamefont {Zhao}, \citenamefont {Hu}, \citenamefont {Gawryluk},
  \citenamefont {Zhu}, \citenamefont {Zhang}, \citenamefont {Meng},
  \citenamefont {Zhen}, \citenamefont {Yu}, \citenamefont {Zhou}, \citenamefont
  {Xu}, \citenamefont {Zhan}, \citenamefont {Pomjakushina},\ and\ \citenamefont
  {Shiroka}}]{Shang2023}%
  \BibitemOpen
  \bibfield  {author} {\bibinfo {author} {\bibfnamefont {T.}~\bibnamefont
  {Shang}}, \bibinfo {author} {\bibfnamefont {J.~Z.}\ \bibnamefont {Zhao}},
  \bibinfo {author} {\bibfnamefont {L.-H.}\ \bibnamefont {Hu}}, \bibinfo
  {author} {\bibfnamefont {D.~J.}\ \bibnamefont {Gawryluk}}, \bibinfo {author}
  {\bibfnamefont {X.~Y.}\ \bibnamefont {Zhu}}, \bibinfo {author} {\bibfnamefont
  {H.}~\bibnamefont {Zhang}}, \bibinfo {author} {\bibfnamefont
  {J.}~\bibnamefont {Meng}}, \bibinfo {author} {\bibfnamefont {Z.~X.}\
  \bibnamefont {Zhen}}, \bibinfo {author} {\bibfnamefont {B.~C.}\ \bibnamefont
  {Yu}}, \bibinfo {author} {\bibfnamefont {Z.}~\bibnamefont {Zhou}}, \bibinfo
  {author} {\bibfnamefont {Y.}~\bibnamefont {Xu}}, \bibinfo {author}
  {\bibfnamefont {Q.~F.}\ \bibnamefont {Zhan}}, \bibinfo {author}
  {\bibfnamefont {E.}~\bibnamefont {Pomjakushina}},\ and\ \bibinfo {author}
  {\bibfnamefont {T.}~\bibnamefont {Shiroka}},\ }\bibfield  {title} {\bibinfo
  {title} {Fully gapped superconductivity and topological aspects of the
  noncentrosymmetric superconductor {TaReSi}},\ }\href
  {https://doi.org/10.1103/PhysRevB.107.224504} {\bibfield  {journal} {\bibinfo
   {journal} {Phys. Rev. B}\ }\textbf {\bibinfo {volume} {107}},\ \bibinfo
  {pages} {224504} (\bibinfo {year} {2023})}\BibitemShut {NoStop}%
\bibitem [{\citenamefont {Xie}\ \emph {et~al.}(2021)\citenamefont {Xie},
  \citenamefont {Gao}, \citenamefont {Xu}, \citenamefont {Zhang}, \citenamefont
  {Hu}, \citenamefont {Gao},\ and\ \citenamefont {Law}}]{Xie2021}%
  \BibitemOpen
  \bibfield  {author} {\bibinfo {author} {\bibfnamefont {Y.-M.}\ \bibnamefont
  {Xie}}, \bibinfo {author} {\bibfnamefont {X.-J.}\ \bibnamefont {Gao}},
  \bibinfo {author} {\bibfnamefont {X.~Y.}\ \bibnamefont {Xu}}, \bibinfo
  {author} {\bibfnamefont {C.-P.}\ \bibnamefont {Zhang}}, \bibinfo {author}
  {\bibfnamefont {J.-X.}\ \bibnamefont {Hu}}, \bibinfo {author} {\bibfnamefont
  {J.~Z.}\ \bibnamefont {Gao}},\ and\ \bibinfo {author} {\bibfnamefont {K.~T.}\
  \bibnamefont {Law}},\ }\bibfield  {title} {\bibinfo {title} {Kramers nodal
  line metals},\ }\href {https://doi.org/10.1038/s41467-021-22903-9} {\bibfield
   {journal} {\bibinfo  {journal} {Nat. Commun.}\ }\textbf {\bibinfo {volume}
  {12}},\ \bibinfo {pages} {3064} (\bibinfo {year} {2021})}\BibitemShut
  {NoStop}%
\bibitem [{\citenamefont {Chang}\ \emph {et~al.}(2018)\citenamefont {Chang},
  \citenamefont {Wieder}, \citenamefont {Schindler}, \citenamefont {Sanchez},
  \citenamefont {Belopolski}, \citenamefont {Huang}, \citenamefont {Singh},
  \citenamefont {Wu}, \citenamefont {Chang}, \citenamefont {Neupert},
  \citenamefont {Xu}, \citenamefont {Lin},\ and\ \citenamefont
  {Hasan}}]{Chang2018}%
  \BibitemOpen
  \bibfield  {author} {\bibinfo {author} {\bibfnamefont {G.}~\bibnamefont
  {Chang}}, \bibinfo {author} {\bibfnamefont {B.~J.}\ \bibnamefont {Wieder}},
  \bibinfo {author} {\bibfnamefont {F.}~\bibnamefont {Schindler}}, \bibinfo
  {author} {\bibfnamefont {D.~S.}\ \bibnamefont {Sanchez}}, \bibinfo {author}
  {\bibfnamefont {I.}~\bibnamefont {Belopolski}}, \bibinfo {author}
  {\bibfnamefont {S.-M.}\ \bibnamefont {Huang}}, \bibinfo {author}
  {\bibfnamefont {B.}~\bibnamefont {Singh}}, \bibinfo {author} {\bibfnamefont
  {D.}~\bibnamefont {Wu}}, \bibinfo {author} {\bibfnamefont {T.-R.}\
  \bibnamefont {Chang}}, \bibinfo {author} {\bibfnamefont {T.}~\bibnamefont
  {Neupert}}, \bibinfo {author} {\bibfnamefont {S.-Y.}\ \bibnamefont {Xu}},
  \bibinfo {author} {\bibfnamefont {H.}~\bibnamefont {Lin}},\ and\ \bibinfo
  {author} {\bibfnamefont {M.~Z.}\ \bibnamefont {Hasan}},\ }\bibfield  {title}
  {\bibinfo {title} {Topological quantum properties of chiral crystals},\
  }\href {https://doi.org/10.1038/s41563-018-0169-3} {\bibfield  {journal}
  {\bibinfo  {journal} {Nat. Mater.}\ }\textbf {\bibinfo {volume} {17}},\
  \bibinfo {pages} {978} (\bibinfo {year} {2018})}\BibitemShut {NoStop}%
\bibitem [{\citenamefont {Bradlyn}\ \emph {et~al.}(2016)\citenamefont
  {Bradlyn}, \citenamefont {Cano}, \citenamefont {Wang}, \citenamefont
  {Vergniory}, \citenamefont {Felser}, \citenamefont {Cava},\ and\
  \citenamefont {Bernevig}}]{Bradlyn2016}%
  \BibitemOpen
  \bibfield  {author} {\bibinfo {author} {\bibfnamefont {B.}~\bibnamefont
  {Bradlyn}}, \bibinfo {author} {\bibfnamefont {J.}~\bibnamefont {Cano}},
  \bibinfo {author} {\bibfnamefont {Z.}~\bibnamefont {Wang}}, \bibinfo {author}
  {\bibfnamefont {M.~G.}\ \bibnamefont {Vergniory}}, \bibinfo {author}
  {\bibfnamefont {C.}~\bibnamefont {Felser}}, \bibinfo {author} {\bibfnamefont
  {R.~J.}\ \bibnamefont {Cava}},\ and\ \bibinfo {author} {\bibfnamefont
  {B.~A.}\ \bibnamefont {Bernevig}},\ }\bibfield  {title} {\bibinfo {title}
  {Beyond {Dirac} and {Weyl} fermions: {Unconventional} quasiparticles in
  conventional crystals},\ }\href {https://doi.org/10.1126/science.aaf5037}
  {\bibfield  {journal} {\bibinfo  {journal} {Science}\ }\textbf {\bibinfo
  {volume} {353}},\ \bibinfo {pages} {aaf5037} (\bibinfo {year}
  {2016})}\BibitemShut {NoStop}%
\bibitem [{\citenamefont {Rao}\ \emph {et~al.}(2019)\citenamefont {Rao},
  \citenamefont {Li}, \citenamefont {Zhang}, \citenamefont {Tian},
  \citenamefont {Li}, \citenamefont {Fu}, \citenamefont {Tang}, \citenamefont
  {Wang}, \citenamefont {Li}, \citenamefont {Fan}, \citenamefont {Li},
  \citenamefont {Huang}, \citenamefont {Liu}, \citenamefont {Long},
  \citenamefont {Fang}, \citenamefont {Weng}, \citenamefont {Shi},
  \citenamefont {Lei}, \citenamefont {Sun}, \citenamefont {Qian},\ and\
  \citenamefont {Ding}}]{Rao2019}%
  \BibitemOpen
  \bibfield  {author} {\bibinfo {author} {\bibfnamefont {Z.}~\bibnamefont
  {Rao}}, \bibinfo {author} {\bibfnamefont {H.}~\bibnamefont {Li}}, \bibinfo
  {author} {\bibfnamefont {T.}~\bibnamefont {Zhang}}, \bibinfo {author}
  {\bibfnamefont {S.}~\bibnamefont {Tian}}, \bibinfo {author} {\bibfnamefont
  {C.}~\bibnamefont {Li}}, \bibinfo {author} {\bibfnamefont {B.}~\bibnamefont
  {Fu}}, \bibinfo {author} {\bibfnamefont {C.}~\bibnamefont {Tang}}, \bibinfo
  {author} {\bibfnamefont {L.}~\bibnamefont {Wang}}, \bibinfo {author}
  {\bibfnamefont {Z.}~\bibnamefont {Li}}, \bibinfo {author} {\bibfnamefont
  {W.}~\bibnamefont {Fan}}, \bibinfo {author} {\bibfnamefont {J.}~\bibnamefont
  {Li}}, \bibinfo {author} {\bibfnamefont {Y.}~\bibnamefont {Huang}}, \bibinfo
  {author} {\bibfnamefont {Z.}~\bibnamefont {Liu}}, \bibinfo {author}
  {\bibfnamefont {Y.}~\bibnamefont {Long}}, \bibinfo {author} {\bibfnamefont
  {C.}~\bibnamefont {Fang}}, \bibinfo {author} {\bibfnamefont {H.}~\bibnamefont
  {Weng}}, \bibinfo {author} {\bibfnamefont {Y.}~\bibnamefont {Shi}}, \bibinfo
  {author} {\bibfnamefont {H.}~\bibnamefont {Lei}}, \bibinfo {author}
  {\bibfnamefont {Y.}~\bibnamefont {Sun}}, \bibinfo {author} {\bibfnamefont
  {T.}~\bibnamefont {Qian}},\ and\ \bibinfo {author} {\bibfnamefont
  {H.}~\bibnamefont {Ding}},\ }\bibfield  {title} {\bibinfo {title}
  {Observation of unconventional chiral fermions with long {Fermi} arcs in
  {CoSi}},\ }\href {https://doi.org/10.1038/s41586-019-1031-8} {\bibfield
  {journal} {\bibinfo  {journal} {Nature}\ }\textbf {\bibinfo {volume} {567}},\
  \bibinfo {pages} {496} (\bibinfo {year} {2019})}\BibitemShut {NoStop}%
\bibitem [{\citenamefont {Sanchez}\ \emph {et~al.}(2019)\citenamefont
  {Sanchez}, \citenamefont {Belopolski}, \citenamefont {Cochran}, \citenamefont
  {Xu}, \citenamefont {Yin}, \citenamefont {Chang}, \citenamefont {Xie},
  \citenamefont {Manna}, \citenamefont {Süß}, \citenamefont {Huang},
  \citenamefont {Alidoust}, \citenamefont {Multer}, \citenamefont {Zhang},
  \citenamefont {Shumiya}, \citenamefont {Wang}, \citenamefont {Wang},
  \citenamefont {Chang}, \citenamefont {Felser}, \citenamefont {Xu},
  \citenamefont {Jia}, \citenamefont {Lin},\ and\ \citenamefont
  {Hasan}}]{Sanchez2019}%
  \BibitemOpen
  \bibfield  {author} {\bibinfo {author} {\bibfnamefont {D.~S.}\ \bibnamefont
  {Sanchez}}, \bibinfo {author} {\bibfnamefont {I.}~\bibnamefont {Belopolski}},
  \bibinfo {author} {\bibfnamefont {T.~A.}\ \bibnamefont {Cochran}}, \bibinfo
  {author} {\bibfnamefont {X.}~\bibnamefont {Xu}}, \bibinfo {author}
  {\bibfnamefont {J.-X.}\ \bibnamefont {Yin}}, \bibinfo {author} {\bibfnamefont
  {G.}~\bibnamefont {Chang}}, \bibinfo {author} {\bibfnamefont
  {W.}~\bibnamefont {Xie}}, \bibinfo {author} {\bibfnamefont {K.}~\bibnamefont
  {Manna}}, \bibinfo {author} {\bibfnamefont {V.}~\bibnamefont {Süß}},
  \bibinfo {author} {\bibfnamefont {C.-Y.}\ \bibnamefont {Huang}}, \bibinfo
  {author} {\bibfnamefont {N.}~\bibnamefont {Alidoust}}, \bibinfo {author}
  {\bibfnamefont {D.}~\bibnamefont {Multer}}, \bibinfo {author} {\bibfnamefont
  {S.~S.}\ \bibnamefont {Zhang}}, \bibinfo {author} {\bibfnamefont
  {N.}~\bibnamefont {Shumiya}}, \bibinfo {author} {\bibfnamefont
  {X.}~\bibnamefont {Wang}}, \bibinfo {author} {\bibfnamefont {G.-Q.}\
  \bibnamefont {Wang}}, \bibinfo {author} {\bibfnamefont {T.-R.}\ \bibnamefont
  {Chang}}, \bibinfo {author} {\bibfnamefont {C.}~\bibnamefont {Felser}},
  \bibinfo {author} {\bibfnamefont {S.-Y.}\ \bibnamefont {Xu}}, \bibinfo
  {author} {\bibfnamefont {S.}~\bibnamefont {Jia}}, \bibinfo {author}
  {\bibfnamefont {H.}~\bibnamefont {Lin}},\ and\ \bibinfo {author}
  {\bibfnamefont {M.~Z.}\ \bibnamefont {Hasan}},\ }\bibfield  {title} {\bibinfo
  {title} {Topological chiral crystals with helicoid-arc quantum states},\
  }\href {https://doi.org/10.1038/s41586-019-1037-2} {\bibfield  {journal}
  {\bibinfo  {journal} {Nature}\ }\textbf {\bibinfo {volume} {567}},\ \bibinfo
  {pages} {500} (\bibinfo {year} {2019})}\BibitemShut {NoStop}%
\bibitem [{\citenamefont {Sato}\ and\ \citenamefont {Ando}(2017)}]{Sato2017}%
  \BibitemOpen
  \bibfield  {author} {\bibinfo {author} {\bibfnamefont {M.}~\bibnamefont
  {Sato}}\ and\ \bibinfo {author} {\bibfnamefont {Y.}~\bibnamefont {Ando}},\
  }\bibfield  {title} {\bibinfo {title} {Topological superconductors: {A}
  review},\ }\href {https://doi.org/10.1088/1361-6633/aa6ac7} {\bibfield
  {journal} {\bibinfo  {journal} {Rep. Prog. Phys.}\ }\textbf {\bibinfo
  {volume} {80}},\ \bibinfo {pages} {076501} (\bibinfo {year}
  {2017})}\BibitemShut {NoStop}%
\bibitem [{\citenamefont {Qi}\ and\ \citenamefont {Zhang}(2011)}]{Qi2011}%
  \BibitemOpen
  \bibfield  {author} {\bibinfo {author} {\bibfnamefont {X.-L.}\ \bibnamefont
  {Qi}}\ and\ \bibinfo {author} {\bibfnamefont {S.-C.}\ \bibnamefont {Zhang}},\
  }\bibfield  {title} {\bibinfo {title} {Topological insulators and
  superconductors},\ }\href {https://doi.org/10.1103/RevModPhys.83.1057}
  {\bibfield  {journal} {\bibinfo  {journal} {Rev. Mod. Phys.}\ }\textbf
  {\bibinfo {volume} {83}},\ \bibinfo {pages} {1057} (\bibinfo {year}
  {2011})}\BibitemShut {NoStop}%
\bibitem [{\citenamefont {Kallin}\ and\ \citenamefont
  {Berlinsky}(2016)}]{Kallin2016}%
  \BibitemOpen
  \bibfield  {author} {\bibinfo {author} {\bibfnamefont {C.}~\bibnamefont
  {Kallin}}\ and\ \bibinfo {author} {\bibfnamefont {J.}~\bibnamefont
  {Berlinsky}},\ }\bibfield  {title} {\bibinfo {title} {Chiral
  superconductors},\ }\href {https://doi.org/10.1088/0034-4885/79/5/054502}
  {\bibfield  {journal} {\bibinfo  {journal} {Rep. Prog. Phys.}\ }\textbf
  {\bibinfo {volume} {79}},\ \bibinfo {pages} {054502} (\bibinfo {year}
  {2016})}\BibitemShut {NoStop}%
\bibitem [{\citenamefont {Guan}\ \emph {et~al.}(2016)\citenamefont {Guan},
  \citenamefont {Chen}, \citenamefont {Chu}, \citenamefont {Sankar},
  \citenamefont {Chou}, \citenamefont {Jeng}, \citenamefont {Chang},\ and\
  \citenamefont {Chuang}}]{Guan2016}%
  \BibitemOpen
  \bibfield  {author} {\bibinfo {author} {\bibfnamefont {S.-Y.}\ \bibnamefont
  {Guan}}, \bibinfo {author} {\bibfnamefont {P.-J.}\ \bibnamefont {Chen}},
  \bibinfo {author} {\bibfnamefont {M.-W.}\ \bibnamefont {Chu}}, \bibinfo
  {author} {\bibfnamefont {R.}~\bibnamefont {Sankar}}, \bibinfo {author}
  {\bibfnamefont {F.}~\bibnamefont {Chou}}, \bibinfo {author} {\bibfnamefont
  {H.-T.}\ \bibnamefont {Jeng}}, \bibinfo {author} {\bibfnamefont {C.-S.}\
  \bibnamefont {Chang}},\ and\ \bibinfo {author} {\bibfnamefont {T.-M.}\
  \bibnamefont {Chuang}},\ }\bibfield  {title} {\bibinfo {title}
  {Superconducting topological surface states in the noncentrosymmetric bulk
  superconductor {PbTaSe$_2$}},\ }\href
  {https://doi.org/10.1126/sciadv.1600894} {\bibfield  {journal} {\bibinfo
  {journal} {Sci. Adv.}\ }\textbf {\bibinfo {volume} {2}},\ \bibinfo {pages}
  {e1600894} (\bibinfo {year} {2016})}\BibitemShut {NoStop}%
\bibitem [{\citenamefont {Sakano}\ \emph {et~al.}(2015)\citenamefont {Sakano},
  \citenamefont {Okawa}, \citenamefont {Kanou}, \citenamefont {Sanjo},
  \citenamefont {Okuda}, \citenamefont {Sasagawa},\ and\ \citenamefont
  {Ishizaka}}]{Sakano2015}%
  \BibitemOpen
  \bibfield  {author} {\bibinfo {author} {\bibfnamefont {M.}~\bibnamefont
  {Sakano}}, \bibinfo {author} {\bibfnamefont {K.}~\bibnamefont {Okawa}},
  \bibinfo {author} {\bibfnamefont {M.}~\bibnamefont {Kanou}}, \bibinfo
  {author} {\bibfnamefont {H.}~\bibnamefont {Sanjo}}, \bibinfo {author}
  {\bibfnamefont {T.}~\bibnamefont {Okuda}}, \bibinfo {author} {\bibfnamefont
  {T.}~\bibnamefont {Sasagawa}},\ and\ \bibinfo {author} {\bibfnamefont
  {K.}~\bibnamefont {Ishizaka}},\ }\bibfield  {title} {\bibinfo {title}
  {Topologically protected surface states in a centrosymmetric superconductor
  $\beta$-{PdBi}$_2$},\ }\href {https://doi.org/10.1038/ncomms9595} {\bibfield
  {journal} {\bibinfo  {journal} {Nat. Commun.}\ }\textbf {\bibinfo {volume}
  {6}},\ \bibinfo {pages} {8595} (\bibinfo {year} {2015})}\BibitemShut
  {NoStop}%
\bibitem [{\citenamefont {Carnicom}\ \emph {et~al.}(2018)\citenamefont
  {Carnicom}, \citenamefont {Xie}, \citenamefont {Klimczuk}, \citenamefont
  {Lin}, \citenamefont {G{\'o}rnicka}, \citenamefont {Sobczak}, \citenamefont
  {Ong},\ and\ \citenamefont {Cava}}]{Carnicom2018}%
  \BibitemOpen
  \bibfield  {author} {\bibinfo {author} {\bibfnamefont {E.~M.}\ \bibnamefont
  {Carnicom}}, \bibinfo {author} {\bibfnamefont {W.}~\bibnamefont {Xie}},
  \bibinfo {author} {\bibfnamefont {T.}~\bibnamefont {Klimczuk}}, \bibinfo
  {author} {\bibfnamefont {J.~J.}\ \bibnamefont {Lin}}, \bibinfo {author}
  {\bibfnamefont {K.}~\bibnamefont {G{\'o}rnicka}}, \bibinfo {author}
  {\bibfnamefont {Z.}~\bibnamefont {Sobczak}}, \bibinfo {author} {\bibfnamefont
  {N.~P.}\ \bibnamefont {Ong}},\ and\ \bibinfo {author} {\bibfnamefont {R.~J.}\
  \bibnamefont {Cava}},\ }\bibfield  {title} {\bibinfo {title}
  {{Ta}{Rh}$_2${B}$_2$ and {Nb}{Rh}$_2${B}$_2$: {S}uperconductors with a chiral
  noncentrosymmetric crystal structure},\ }\href
  {https://doi.org/10.1126/sciadv.aar7969} {\bibfield  {journal} {\bibinfo
  {journal} {Sci. Adv.}\ }\textbf {\bibinfo {volume} {4}},\ \bibinfo {pages}
  {eaar7969} (\bibinfo {year} {2018})}\BibitemShut {NoStop}%
\bibitem [{\citenamefont {Bauer}\ \emph {et~al.}(2004)\citenamefont {Bauer},
  \citenamefont {Hilscher}, \citenamefont {Michor}, \citenamefont {Paul},
  \citenamefont {Scheidt}, \citenamefont {Gribanov}, \citenamefont {Seropegin},
  \citenamefont {No{\"e}l}, \citenamefont {Sigrist},\ and\ \citenamefont
  {Rogl}}]{Bauer2004}%
  \BibitemOpen
  \bibfield  {author} {\bibinfo {author} {\bibfnamefont {E.}~\bibnamefont
  {Bauer}}, \bibinfo {author} {\bibfnamefont {G.}~\bibnamefont {Hilscher}},
  \bibinfo {author} {\bibfnamefont {H.}~\bibnamefont {Michor}}, \bibinfo
  {author} {\bibfnamefont {C.}~\bibnamefont {Paul}}, \bibinfo {author}
  {\bibfnamefont {E.~W.}\ \bibnamefont {Scheidt}}, \bibinfo {author}
  {\bibfnamefont {A.}~\bibnamefont {Gribanov}}, \bibinfo {author}
  {\bibfnamefont {Y.}~\bibnamefont {Seropegin}}, \bibinfo {author}
  {\bibfnamefont {H.}~\bibnamefont {No{\"e}l}}, \bibinfo {author}
  {\bibfnamefont {M.}~\bibnamefont {Sigrist}},\ and\ \bibinfo {author}
  {\bibfnamefont {P.}~\bibnamefont {Rogl}},\ }\bibfield  {title} {\bibinfo
  {title} {Heavy fermion superconductivity and magnetic order in
  noncentrosymmetric {Ce}{Pt}$_{3}${Si}},\ }\href
  {https://doi.org/10.1103/PhysRevLett.92.027003} {\bibfield  {journal}
  {\bibinfo  {journal} {Phys. Rev. Lett.}\ }\textbf {\bibinfo {volume} {92}},\
  \bibinfo {pages} {027003} (\bibinfo {year} {2004})}\BibitemShut {NoStop}%
\bibitem [{\citenamefont {Su}\ \emph {et~al.}(2021)\citenamefont {Su},
  \citenamefont {Shang}, \citenamefont {Du}, \citenamefont {Chen},
  \citenamefont {Ye}, \citenamefont {Lu}, \citenamefont {Cao}, \citenamefont
  {Smidman},\ and\ \citenamefont {Yuan}}]{Su2021a}%
  \BibitemOpen
  \bibfield  {author} {\bibinfo {author} {\bibfnamefont {H.}~\bibnamefont
  {Su}}, \bibinfo {author} {\bibfnamefont {T.}~\bibnamefont {Shang}}, \bibinfo
  {author} {\bibfnamefont {F.}~\bibnamefont {Du}}, \bibinfo {author}
  {\bibfnamefont {C.~F.}\ \bibnamefont {Chen}}, \bibinfo {author}
  {\bibfnamefont {H.~Q.}\ \bibnamefont {Ye}}, \bibinfo {author} {\bibfnamefont
  {X.}~\bibnamefont {Lu}}, \bibinfo {author} {\bibfnamefont {C.}~\bibnamefont
  {Cao}}, \bibinfo {author} {\bibfnamefont {M.}~\bibnamefont {Smidman}},\ and\
  \bibinfo {author} {\bibfnamefont {H.~Q.}\ \bibnamefont {Yuan}},\ }\bibfield
  {title} {\bibinfo {title} {{NbReSi}: A noncentrosymetric superconductor with
  large upper critical field},\ }\href
  {https://doi.org/10.1103/PhysRevMaterials.5.114802} {\bibfield  {journal}
  {\bibinfo  {journal} {Phys. Rev. Materials}\ }\textbf {\bibinfo {volume}
  {5}},\ \bibinfo {pages} {114802} (\bibinfo {year} {2021})}\BibitemShut
  {NoStop}%
\bibitem [{\citenamefont {Yuan}\ \emph {et~al.}(2006)\citenamefont {Yuan},
  \citenamefont {Agterberg}, \citenamefont {Hayashi}, \citenamefont {Badica},
  \citenamefont {Vandervelde}, \citenamefont {Togano}, \citenamefont
  {Sigrist},\ and\ \citenamefont {Salamon}}]{yuan2006}%
  \BibitemOpen
  \bibfield  {author} {\bibinfo {author} {\bibfnamefont {H.~Q.}\ \bibnamefont
  {Yuan}}, \bibinfo {author} {\bibfnamefont {D.~F.}\ \bibnamefont {Agterberg}},
  \bibinfo {author} {\bibfnamefont {N.}~\bibnamefont {Hayashi}}, \bibinfo
  {author} {\bibfnamefont {P.}~\bibnamefont {Badica}}, \bibinfo {author}
  {\bibfnamefont {D.}~\bibnamefont {Vandervelde}}, \bibinfo {author}
  {\bibfnamefont {K.}~\bibnamefont {Togano}}, \bibinfo {author} {\bibfnamefont
  {M.}~\bibnamefont {Sigrist}},\ and\ \bibinfo {author} {\bibfnamefont {M.~B.}\
  \bibnamefont {Salamon}},\ }\bibfield  {title} {\bibinfo {title} {{$s$}-wave
  spin-triplet order in superconductors without inversion symmetry:
  {Li}$_{2}${Pd}$_{3}${B} and {Li}$_{2}${Pt}$_{3}${B}},\ }\href
  {https://doi.org/10.1103/PhysRevLett.97.017006} {\bibfield  {journal}
  {\bibinfo  {journal} {Phys. Rev. Lett.}\ }\textbf {\bibinfo {volume} {97}},\
  \bibinfo {pages} {017006} (\bibinfo {year} {2006})}\BibitemShut {NoStop}%
\bibitem [{\citenamefont {Nishiyama}\ \emph {et~al.}(2007)\citenamefont
  {Nishiyama}, \citenamefont {Inada},\ and\ \citenamefont
  {Zheng}}]{nishiyama2007}%
  \BibitemOpen
  \bibfield  {author} {\bibinfo {author} {\bibfnamefont {M.}~\bibnamefont
  {Nishiyama}}, \bibinfo {author} {\bibfnamefont {Y.}~\bibnamefont {Inada}},\
  and\ \bibinfo {author} {\bibfnamefont {G.-q.}\ \bibnamefont {Zheng}},\
  }\bibfield  {title} {\bibinfo {title} {Spin triplet superconducting state due
  to broken inversion symmetry in {Li}$_{2}${Pt}$_{3}${B}},\ }\href
  {https://doi.org/10.1103/PhysRevLett.98.047002} {\bibfield  {journal}
  {\bibinfo  {journal} {Phys. Rev. Lett.}\ }\textbf {\bibinfo {volume} {98}},\
  \bibinfo {pages} {047002} (\bibinfo {year} {2007})}\BibitemShut {NoStop}%
\bibitem [{\citenamefont {Bonalde}\ \emph {et~al.}(2005)\citenamefont
  {Bonalde}, \citenamefont {Br{\"a}mer-Escamilla},\ and\ \citenamefont
  {Bauer}}]{bonalde2005CePt3Si}%
  \BibitemOpen
  \bibfield  {author} {\bibinfo {author} {\bibfnamefont {I.}~\bibnamefont
  {Bonalde}}, \bibinfo {author} {\bibfnamefont {W.}~\bibnamefont
  {Br{\"a}mer-Escamilla}},\ and\ \bibinfo {author} {\bibfnamefont
  {E.}~\bibnamefont {Bauer}},\ }\bibfield  {title} {\bibinfo {title} {Evidence
  for line nodes in the superconducting energy gap of noncentrosymmetric
  {Ce}{Pt}$_{3}${Si} from magnetic penetration depth measurements},\ }\href
  {https://doi.org/10.1103/PhysRevLett.94.207002} {\bibfield  {journal}
  {\bibinfo  {journal} {Phys. Rev. Lett.}\ }\textbf {\bibinfo {volume} {94}},\
  \bibinfo {pages} {207002} (\bibinfo {year} {2005})}\BibitemShut {NoStop}%
\bibitem [{\citenamefont {Shang}\ \emph
  {et~al.}(2020{\natexlab{a}})\citenamefont {Shang}, \citenamefont {Smidman},
  \citenamefont {Wang}, \citenamefont {Chang}, \citenamefont {Baines},
  \citenamefont {Lee}, \citenamefont {Nie}, \citenamefont {Pang}, \citenamefont
  {Xie}, \citenamefont {Jiang}, \citenamefont {Shi}, \citenamefont {Medarde},
  \citenamefont {Shiroka},\ and\ \citenamefont {Yuan}}]{Shang2020}%
  \BibitemOpen
  \bibfield  {author} {\bibinfo {author} {\bibfnamefont {T.}~\bibnamefont
  {Shang}}, \bibinfo {author} {\bibfnamefont {M.}~\bibnamefont {Smidman}},
  \bibinfo {author} {\bibfnamefont {A.}~\bibnamefont {Wang}}, \bibinfo {author}
  {\bibfnamefont {L.-J.}\ \bibnamefont {Chang}}, \bibinfo {author}
  {\bibfnamefont {C.}~\bibnamefont {Baines}}, \bibinfo {author} {\bibfnamefont
  {M.~K.}\ \bibnamefont {Lee}}, \bibinfo {author} {\bibfnamefont {Z.~Y.}\
  \bibnamefont {Nie}}, \bibinfo {author} {\bibfnamefont {G.~M.}\ \bibnamefont
  {Pang}}, \bibinfo {author} {\bibfnamefont {W.}~\bibnamefont {Xie}}, \bibinfo
  {author} {\bibfnamefont {W.~B.}\ \bibnamefont {Jiang}}, \bibinfo {author}
  {\bibfnamefont {M.}~\bibnamefont {Shi}}, \bibinfo {author} {\bibfnamefont
  {M.}~\bibnamefont {Medarde}}, \bibinfo {author} {\bibfnamefont
  {T.}~\bibnamefont {Shiroka}},\ and\ \bibinfo {author} {\bibfnamefont {H.~Q.}\
  \bibnamefont {Yuan}},\ }\bibfield  {title} {\bibinfo {title} {Simultaneous
  nodal superconductivity and time-reversal symmetry breaking in the
  noncentrosymmetric superconductor {CaPtAs}},\ }\href
  {https://doi.org/10.1103/PhysRevLett.124.207001} {\bibfield  {journal}
  {\bibinfo  {journal} {Phys. Rev. Lett.}\ }\textbf {\bibinfo {volume} {124}},\
  \bibinfo {pages} {207001} (\bibinfo {year} {2020}{\natexlab{a}})}\BibitemShut
  {NoStop}%
\bibitem [{\citenamefont {Kuroiwa}\ \emph {et~al.}(2008)\citenamefont
  {Kuroiwa}, \citenamefont {Saura}, \citenamefont {Akimitsu}, \citenamefont
  {Hiraishi}, \citenamefont {Miyazaki}, \citenamefont {Satoh}, \citenamefont
  {Takeshita},\ and\ \citenamefont {Kadono}}]{kuroiwa2008}%
  \BibitemOpen
  \bibfield  {author} {\bibinfo {author} {\bibfnamefont {S.}~\bibnamefont
  {Kuroiwa}}, \bibinfo {author} {\bibfnamefont {Y.}~\bibnamefont {Saura}},
  \bibinfo {author} {\bibfnamefont {J.}~\bibnamefont {Akimitsu}}, \bibinfo
  {author} {\bibfnamefont {M.}~\bibnamefont {Hiraishi}}, \bibinfo {author}
  {\bibfnamefont {M.}~\bibnamefont {Miyazaki}}, \bibinfo {author}
  {\bibfnamefont {K.~H.}\ \bibnamefont {Satoh}}, \bibinfo {author}
  {\bibfnamefont {S.}~\bibnamefont {Takeshita}},\ and\ \bibinfo {author}
  {\bibfnamefont {R.}~\bibnamefont {Kadono}},\ }\bibfield  {title} {\bibinfo
  {title} {Multigap superconductivity in sesquicarbides {La}$_{2}${C}$_{3}$ and
  {Y}$_{2}${C}$_{3}$},\ }\href {https://doi.org/10.1103/PhysRevLett.100.097002}
  {\bibfield  {journal} {\bibinfo  {journal} {Phys. Rev. Lett.}\ }\textbf
  {\bibinfo {volume} {100}},\ \bibinfo {pages} {097002} (\bibinfo {year}
  {2008})}\BibitemShut {NoStop}%
\bibitem [{\citenamefont {Sundar}\ \emph {et~al.}(2021)\citenamefont {Sundar},
  \citenamefont {Dunsiger}, \citenamefont {Gheidi}, \citenamefont {Akella},
  \citenamefont {C\^ot\'e}, \citenamefont {\"Ozdemir}, \citenamefont
  {Lee-Hone}, \citenamefont {Broun}, \citenamefont {Mun}, \citenamefont
  {Honda}, \citenamefont {Sato}, \citenamefont {Koizumi}, \citenamefont
  {Settai}, \citenamefont {Hirose}, \citenamefont {Bonalde},\ and\
  \citenamefont {Sonier}}]{Sundar2021}%
  \BibitemOpen
  \bibfield  {author} {\bibinfo {author} {\bibfnamefont {S.}~\bibnamefont
  {Sundar}}, \bibinfo {author} {\bibfnamefont {S.~R.}\ \bibnamefont
  {Dunsiger}}, \bibinfo {author} {\bibfnamefont {S.}~\bibnamefont {Gheidi}},
  \bibinfo {author} {\bibfnamefont {K.~S.}\ \bibnamefont {Akella}}, \bibinfo
  {author} {\bibfnamefont {A.~M.}\ \bibnamefont {C\^ot\'e}}, \bibinfo {author}
  {\bibfnamefont {H.~U.}\ \bibnamefont {\"Ozdemir}}, \bibinfo {author}
  {\bibfnamefont {N.~R.}\ \bibnamefont {Lee-Hone}}, \bibinfo {author}
  {\bibfnamefont {D.~M.}\ \bibnamefont {Broun}}, \bibinfo {author}
  {\bibfnamefont {E.}~\bibnamefont {Mun}}, \bibinfo {author} {\bibfnamefont
  {F.}~\bibnamefont {Honda}}, \bibinfo {author} {\bibfnamefont {Y.~J.}\
  \bibnamefont {Sato}}, \bibinfo {author} {\bibfnamefont {T.}~\bibnamefont
  {Koizumi}}, \bibinfo {author} {\bibfnamefont {R.}~\bibnamefont {Settai}},
  \bibinfo {author} {\bibfnamefont {Y.}~\bibnamefont {Hirose}}, \bibinfo
  {author} {\bibfnamefont {I.}~\bibnamefont {Bonalde}},\ and\ \bibinfo {author}
  {\bibfnamefont {J.~E.}\ \bibnamefont {Sonier}},\ }\bibfield  {title}
  {\bibinfo {title} {Two-gap time reversal symmetry breaking superconductivity
  in noncentrosymmetric {LaNiC}$_{2}$},\ }\href
  {https://doi.org/10.1103/PhysRevB.103.014511} {\bibfield  {journal} {\bibinfo
   {journal} {Phys. Rev. B}\ }\textbf {\bibinfo {volume} {103}},\ \bibinfo
  {pages} {014511} (\bibinfo {year} {2021})}\BibitemShut {NoStop}%
\bibitem [{\citenamefont {Hillier}\ \emph {et~al.}(2009)\citenamefont
  {Hillier}, \citenamefont {Quintanilla},\ and\ \citenamefont
  {Cywinski}}]{Hillier2009}%
  \BibitemOpen
  \bibfield  {author} {\bibinfo {author} {\bibfnamefont {A.~D.}\ \bibnamefont
  {Hillier}}, \bibinfo {author} {\bibfnamefont {J.}~\bibnamefont
  {Quintanilla}},\ and\ \bibinfo {author} {\bibfnamefont {R.}~\bibnamefont
  {Cywinski}},\ }\bibfield  {title} {\bibinfo {title} {Evidence for
  time-reversal symmetry breaking in the noncentrosymmetric superconductor
  {LaNiC$_{2}$}},\ }\href {https://doi.org/10.1103/PhysRevLett.102.117007}
  {\bibfield  {journal} {\bibinfo  {journal} {Phys. Rev. Lett.}\ }\textbf
  {\bibinfo {volume} {102}},\ \bibinfo {pages} {117007} (\bibinfo {year}
  {2009})}\BibitemShut {NoStop}%
\bibitem [{\citenamefont {Barker}\ \emph {et~al.}(2015)\citenamefont {Barker},
  \citenamefont {Singh}, \citenamefont {Thamizhavel}, \citenamefont {Hillier},
  \citenamefont {Lees}, \citenamefont {Balakrishnan}, \citenamefont {Paul},\
  and\ \citenamefont {Singh}}]{Barker2015}%
  \BibitemOpen
  \bibfield  {author} {\bibinfo {author} {\bibfnamefont {J.~A.~T.}\
  \bibnamefont {Barker}}, \bibinfo {author} {\bibfnamefont {D.}~\bibnamefont
  {Singh}}, \bibinfo {author} {\bibfnamefont {A.}~\bibnamefont {Thamizhavel}},
  \bibinfo {author} {\bibfnamefont {A.~D.}\ \bibnamefont {Hillier}}, \bibinfo
  {author} {\bibfnamefont {M.~R.}\ \bibnamefont {Lees}}, \bibinfo {author}
  {\bibfnamefont {G.}~\bibnamefont {Balakrishnan}}, \bibinfo {author}
  {\bibfnamefont {D.~M.}\ \bibnamefont {Paul}},\ and\ \bibinfo {author}
  {\bibfnamefont {R.~P.}\ \bibnamefont {Singh}},\ }\bibfield  {title} {\bibinfo
  {title} {Unconventional superconductivity in {La}$_{7}${Ir}$_{3}$ revealed by
  muon spin relaxation: Introducing a new family of noncentrosymmetric
  superconductor that breaks time-reversal symmetry},\ }\href
  {https://doi.org/10.1103/PhysRevLett.115.267001} {\bibfield  {journal}
  {\bibinfo  {journal} {Phys. Rev. Lett.}\ }\textbf {\bibinfo {volume} {115}},\
  \bibinfo {pages} {267001} (\bibinfo {year} {2015})}\BibitemShut {NoStop}%
\bibitem [{\citenamefont {Shang}\ \emph
  {et~al.}(2020{\natexlab{b}})\citenamefont {Shang}, \citenamefont {Ghosh},
  \citenamefont {Zhao}, \citenamefont {Chang}, \citenamefont {Baines},
  \citenamefont {Lee}, \citenamefont {Gawryluk}, \citenamefont {Shi},
  \citenamefont {Medarde}, \citenamefont {Quintanilla},\ and\ \citenamefont
  {Shiroka}}]{Shang2020b}%
  \BibitemOpen
  \bibfield  {author} {\bibinfo {author} {\bibfnamefont {T.}~\bibnamefont
  {Shang}}, \bibinfo {author} {\bibfnamefont {S.~K.}\ \bibnamefont {Ghosh}},
  \bibinfo {author} {\bibfnamefont {J.~Z.}\ \bibnamefont {Zhao}}, \bibinfo
  {author} {\bibfnamefont {L.-J.}\ \bibnamefont {Chang}}, \bibinfo {author}
  {\bibfnamefont {C.}~\bibnamefont {Baines}}, \bibinfo {author} {\bibfnamefont
  {M.~K.}\ \bibnamefont {Lee}}, \bibinfo {author} {\bibfnamefont {D.~J.}\
  \bibnamefont {Gawryluk}}, \bibinfo {author} {\bibfnamefont {M.}~\bibnamefont
  {Shi}}, \bibinfo {author} {\bibfnamefont {M.}~\bibnamefont {Medarde}},
  \bibinfo {author} {\bibfnamefont {J.}~\bibnamefont {Quintanilla}},\ and\
  \bibinfo {author} {\bibfnamefont {T.}~\bibnamefont {Shiroka}},\ }\bibfield
  {title} {\bibinfo {title} {Time-reversal symmetry breaking in the
  noncentrosymmetric {Zr$_3$Ir} superconductor},\ }\href
  {https://doi.org/10.1103/PhysRevB.102.020503} {\bibfield  {journal} {\bibinfo
   {journal} {Phys. Rev. B}\ }\textbf {\bibinfo {volume} {102}},\ \bibinfo
  {pages} {020503(R)} (\bibinfo {year} {2020}{\natexlab{b}})}\BibitemShut
  {NoStop}%
\bibitem [{\citenamefont {Singh}\ \emph {et~al.}(2014)\citenamefont {Singh},
  \citenamefont {Hillier}, \citenamefont {Mazidian}, \citenamefont
  {Quintanilla}, \citenamefont {Annett}, \citenamefont {Paul}, \citenamefont
  {Balakrishnan},\ and\ \citenamefont {Lees}}]{Singh2014}%
  \BibitemOpen
  \bibfield  {author} {\bibinfo {author} {\bibfnamefont {R.~P.}\ \bibnamefont
  {Singh}}, \bibinfo {author} {\bibfnamefont {A.~D.}\ \bibnamefont {Hillier}},
  \bibinfo {author} {\bibfnamefont {B.}~\bibnamefont {Mazidian}}, \bibinfo
  {author} {\bibfnamefont {J.}~\bibnamefont {Quintanilla}}, \bibinfo {author}
  {\bibfnamefont {J.~F.}\ \bibnamefont {Annett}}, \bibinfo {author}
  {\bibfnamefont {D.~M.}\ \bibnamefont {Paul}}, \bibinfo {author}
  {\bibfnamefont {G.}~\bibnamefont {Balakrishnan}},\ and\ \bibinfo {author}
  {\bibfnamefont {M.~R.}\ \bibnamefont {Lees}},\ }\bibfield  {title} {\bibinfo
  {title} {Detection of time-reversal symmetry breaking in the
  noncentrosymmetric superconductor {Re}$_{6}${Zr} using muon-spin
  spectroscopy},\ }\href {https://doi.org/10.1103/PhysRevLett.112.107002}
  {\bibfield  {journal} {\bibinfo  {journal} {Phys. Rev. Lett.}\ }\textbf
  {\bibinfo {volume} {112}},\ \bibinfo {pages} {107002} (\bibinfo {year}
  {2014})}\BibitemShut {NoStop}%
\bibitem [{\citenamefont {Shang}\ \emph {et~al.}(2018)\citenamefont {Shang},
  \citenamefont {Smidman}, \citenamefont {Ghosh}, \citenamefont {Baines},
  \citenamefont {Chang}, \citenamefont {Gawryluk}, \citenamefont {Barker},
  \citenamefont {Singh}, \citenamefont {Paul}, \citenamefont {Balakrishnan},
  \citenamefont {Pomjakushina}, \citenamefont {Shi}, \citenamefont {Medarde},
  \citenamefont {Hillier}, \citenamefont {Yuan}, \citenamefont {Quintanilla},
  \citenamefont {Mesot},\ and\ \citenamefont {Shiroka}}]{Shang2018b}%
  \BibitemOpen
  \bibfield  {author} {\bibinfo {author} {\bibfnamefont {T.}~\bibnamefont
  {Shang}}, \bibinfo {author} {\bibfnamefont {M.}~\bibnamefont {Smidman}},
  \bibinfo {author} {\bibfnamefont {S.~K.}\ \bibnamefont {Ghosh}}, \bibinfo
  {author} {\bibfnamefont {C.}~\bibnamefont {Baines}}, \bibinfo {author}
  {\bibfnamefont {L.~J.}\ \bibnamefont {Chang}}, \bibinfo {author}
  {\bibfnamefont {D.~J.}\ \bibnamefont {Gawryluk}}, \bibinfo {author}
  {\bibfnamefont {J.~A.~T.}\ \bibnamefont {Barker}}, \bibinfo {author}
  {\bibfnamefont {R.~P.}\ \bibnamefont {Singh}}, \bibinfo {author}
  {\bibfnamefont {D.~M.}\ \bibnamefont {Paul}}, \bibinfo {author}
  {\bibfnamefont {G.}~\bibnamefont {Balakrishnan}}, \bibinfo {author}
  {\bibfnamefont {E.}~\bibnamefont {Pomjakushina}}, \bibinfo {author}
  {\bibfnamefont {M.}~\bibnamefont {Shi}}, \bibinfo {author} {\bibfnamefont
  {M.}~\bibnamefont {Medarde}}, \bibinfo {author} {\bibfnamefont {A.~D.}\
  \bibnamefont {Hillier}}, \bibinfo {author} {\bibfnamefont {H.~Q.}\
  \bibnamefont {Yuan}}, \bibinfo {author} {\bibfnamefont {J.}~\bibnamefont
  {Quintanilla}}, \bibinfo {author} {\bibfnamefont {J.}~\bibnamefont {Mesot}},\
  and\ \bibinfo {author} {\bibfnamefont {T.}~\bibnamefont {Shiroka}},\
  }\bibfield  {title} {\bibinfo {title} {Time-reversal symmetry breaking in
  {Re}-based superconductors},\ }\href
  {https://doi.org/10.1103/PhysRevLett.121.257002} {\bibfield  {journal}
  {\bibinfo  {journal} {Phys. Rev. Lett.}\ }\textbf {\bibinfo {volume} {121}},\
  \bibinfo {pages} {257002} (\bibinfo {year} {2018})}\BibitemShut {NoStop}%
\bibitem [{\citenamefont {Welter}\ \emph {et~al.}(1993)\citenamefont {Welter},
  \citenamefont {Venturini}, \citenamefont {Malaman},\ and\ \citenamefont
  {Ressouche}}]{Welter1993}%
  \BibitemOpen
  \bibfield  {author} {\bibinfo {author} {\bibfnamefont {R.}~\bibnamefont
  {Welter}}, \bibinfo {author} {\bibfnamefont {G.}~\bibnamefont {Venturini}},
  \bibinfo {author} {\bibfnamefont {B.}~\bibnamefont {Malaman}},\ and\ \bibinfo
  {author} {\bibfnamefont {E.}~\bibnamefont {Ressouche}},\ }\bibfield  {title}
  {\bibinfo {title} {Crystallographic data and magnetic properties of new {RTX}
  compounds ({R} $\equiv$ {La}-{Sm}, {Gd}; {T} $\equiv$ {Ru}, {Os}; {X}
  $\equiv$ {Si}, {Ge}). {Magnetic} structure of {NdRuSi}},\ }\href
  {https://doi.org/10.1016/0925-8388(93)90536-V} {\bibfield  {journal}
  {\bibinfo  {journal} {J. Alloys Compd.}\ }\textbf {\bibinfo {volume} {202}},\
  \bibinfo {pages} {165} (\bibinfo {year} {1993})}\BibitemShut {NoStop}%
\bibitem [{\citenamefont {Morozkin}\ \emph {et~al.}(1999)\citenamefont
  {Morozkin}, \citenamefont {Seropegin}, \citenamefont {Sviridov},\ and\
  \citenamefont {Riabinkin}}]{Morozkin1999}%
  \BibitemOpen
  \bibfield  {author} {\bibinfo {author} {\bibfnamefont {A.~V.}\ \bibnamefont
  {Morozkin}}, \bibinfo {author} {\bibfnamefont {Y.~D.}\ \bibnamefont
  {Seropegin}}, \bibinfo {author} {\bibfnamefont {I.~A.}\ \bibnamefont
  {Sviridov}},\ and\ \bibinfo {author} {\bibfnamefont {I.~G.}\ \bibnamefont
  {Riabinkin}},\ }\bibfield  {title} {\bibinfo {title} {Crystallographic data
  of new ternary {Co$_2$Si}-type {RTSi} ({R} = {Y}, {Tb}-{Tm}, {T} = {Mn},
  {Ru}) compounds},\ }\href {https://doi.org/10.1016/S0925-8388(98)00784-1}
  {\bibfield  {journal} {\bibinfo  {journal} {J. Alloys Compd.}\ }\textbf
  {\bibinfo {volume} {282}},\ \bibinfo {pages} {L4} (\bibinfo {year}
  {1999})}\BibitemShut {NoStop}%
\bibitem [{\citenamefont {Subba~Rao}\ \emph {et~al.}(1985)\citenamefont
  {Subba~Rao}, \citenamefont {Wagner}, \citenamefont {Balakrishnan},
  \citenamefont {Janaki}, \citenamefont {Paulus}, \citenamefont {Sch\"ollhorn},
  \citenamefont {Subramanian},\ and\ \citenamefont {Poppe}}]{Subba1985}%
  \BibitemOpen
  \bibfield  {author} {\bibinfo {author} {\bibfnamefont {G.~V.}\ \bibnamefont
  {Subba~Rao}}, \bibinfo {author} {\bibfnamefont {K.}~\bibnamefont {Wagner}},
  \bibinfo {author} {\bibfnamefont {G.}~\bibnamefont {Balakrishnan}}, \bibinfo
  {author} {\bibfnamefont {J.}~\bibnamefont {Janaki}}, \bibinfo {author}
  {\bibfnamefont {W.}~\bibnamefont {Paulus}}, \bibinfo {author} {\bibfnamefont
  {R.}~\bibnamefont {Sch\"ollhorn}}, \bibinfo {author} {\bibfnamefont {V.~S.}\
  \bibnamefont {Subramanian}},\ and\ \bibinfo {author} {\bibfnamefont
  {U.}~\bibnamefont {Poppe}},\ }\bibfield  {title} {\bibinfo {title} {Structure
  and superconductivity studies on ternary equiatomic silicides {$MM$}'{Si}},\
  }\href {https://doi.org/10.1007/BF02747575} {\bibfield  {journal} {\bibinfo
  {journal} {Bull. Mater. Sci.}\ }\textbf {\bibinfo {volume} {7}},\ \bibinfo
  {pages} {215} (\bibinfo {year} {1985})}\BibitemShut {NoStop}%
\bibitem [{\citenamefont {Meisner}\ \emph {et~al.}(1983)\citenamefont
  {Meisner}, \citenamefont {Ku},\ and\ \citenamefont {Barz}}]{Meisner1983}%
  \BibitemOpen
  \bibfield  {author} {\bibinfo {author} {\bibfnamefont {G.~P.}\ \bibnamefont
  {Meisner}}, \bibinfo {author} {\bibfnamefont {H.~C.}\ \bibnamefont {Ku}},\
  and\ \bibinfo {author} {\bibfnamefont {H.}~\bibnamefont {Barz}},\ }\bibfield
  {title} {\bibinfo {title} {Superconducting equiatomic ternary transition
  metal arsenides},\ }\href
  {https://doi.org/https://doi.org/10.1016/0025-5408(83)90010-7} {\bibfield
  {journal} {\bibinfo  {journal} {Mater. Res. Bull.}\ }\textbf {\bibinfo
  {volume} {18}},\ \bibinfo {pages} {983} (\bibinfo {year} {1983})}\BibitemShut
  {NoStop}%
\bibitem [{\citenamefont {Shirotani}\ \emph {et~al.}(1999)\citenamefont
  {Shirotani}, \citenamefont {Tachi}, \citenamefont {Konno}, \citenamefont
  {Todo},\ and\ \citenamefont {Yagi}}]{Ichimin1999}%
  \BibitemOpen
  \bibfield  {author} {\bibinfo {author} {\bibfnamefont {I.}~\bibnamefont
  {Shirotani}}, \bibinfo {author} {\bibfnamefont {K.}~\bibnamefont {Tachi}},
  \bibinfo {author} {\bibfnamefont {Y.}~\bibnamefont {Konno}}, \bibinfo
  {author} {\bibfnamefont {S.}~\bibnamefont {Todo}},\ and\ \bibinfo {author}
  {\bibfnamefont {T.}~\bibnamefont {Yagi}},\ }\bibfield  {title} {\bibinfo
  {title} {Superconductivity of the ternary ruthenium compounds {HfRuP} and
  {ZrRuX} ({X} = {P}, {As}, {Si} or {Ge}) prepared at a high pressure},\ }\href
  {https://doi.org/10.1080/13642819908205748} {\bibfield  {journal} {\bibinfo
  {journal} {Philos. Mag. B}\ }\textbf {\bibinfo {volume} {79}},\ \bibinfo
  {pages} {767} (\bibinfo {year} {1999})}\BibitemShut {NoStop}%
\bibitem [{\citenamefont {Das}\ \emph {et~al.}(2021)\citenamefont {Das},
  \citenamefont {Adroja}, \citenamefont {Lees}, \citenamefont {Taylor},
  \citenamefont {Bishnoi}, \citenamefont {Anand}, \citenamefont
  {Bhattacharyya}, \citenamefont {Guguchia}, \citenamefont {Baines},
  \citenamefont {Luetkens}, \citenamefont {Stenning}, \citenamefont {Duan},
  \citenamefont {Wang},\ and\ \citenamefont {Jin}}]{Das2021}%
  \BibitemOpen
  \bibfield  {author} {\bibinfo {author} {\bibfnamefont {D.}~\bibnamefont
  {Das}}, \bibinfo {author} {\bibfnamefont {D.~T.}\ \bibnamefont {Adroja}},
  \bibinfo {author} {\bibfnamefont {M.~R.}\ \bibnamefont {Lees}}, \bibinfo
  {author} {\bibfnamefont {R.~W.}\ \bibnamefont {Taylor}}, \bibinfo {author}
  {\bibfnamefont {Z.~S.}\ \bibnamefont {Bishnoi}}, \bibinfo {author}
  {\bibfnamefont {V.~K.}\ \bibnamefont {Anand}}, \bibinfo {author}
  {\bibfnamefont {A.}~\bibnamefont {Bhattacharyya}}, \bibinfo {author}
  {\bibfnamefont {Z.}~\bibnamefont {Guguchia}}, \bibinfo {author}
  {\bibfnamefont {C.}~\bibnamefont {Baines}}, \bibinfo {author} {\bibfnamefont
  {H.}~\bibnamefont {Luetkens}}, \bibinfo {author} {\bibfnamefont {G.~B.~G.}\
  \bibnamefont {Stenning}}, \bibinfo {author} {\bibfnamefont {L.}~\bibnamefont
  {Duan}}, \bibinfo {author} {\bibfnamefont {X.}~\bibnamefont {Wang}},\ and\
  \bibinfo {author} {\bibfnamefont {C.}~\bibnamefont {Jin}},\ }\bibfield
  {title} {\bibinfo {title} {Probing the superconducting gap structure in the
  noncentrosymmetric topological superconductor {ZrRuAs}},\ }\href
  {https://doi.org/10.1103/PhysRevB.103.144516} {\bibfield  {journal} {\bibinfo
   {journal} {Phys. Rev. B}\ }\textbf {\bibinfo {volume} {103}},\ \bibinfo
  {pages} {144516} (\bibinfo {year} {2021})}\BibitemShut {NoStop}%
\bibitem [{\citenamefont {Das}\ \emph {et~al.}(2023)\citenamefont {Das},
  \citenamefont {Adroja}, \citenamefont {Tripathi}, \citenamefont {Guguchia},
  \citenamefont {Hotz}, \citenamefont {Luetkens}, \citenamefont {Wang},
  \citenamefont {Yan}, \citenamefont {Luo},\ and\ \citenamefont
  {Shi}}]{Das2023}%
  \BibitemOpen
  \bibfield  {author} {\bibinfo {author} {\bibfnamefont {D.}~\bibnamefont
  {Das}}, \bibinfo {author} {\bibfnamefont {D.}~\bibnamefont {Adroja}},
  \bibinfo {author} {\bibfnamefont {R.}~\bibnamefont {Tripathi}}, \bibinfo
  {author} {\bibfnamefont {Z.}~\bibnamefont {Guguchia}}, \bibinfo {author}
  {\bibfnamefont {F.}~\bibnamefont {Hotz}}, \bibinfo {author} {\bibfnamefont
  {H.}~\bibnamefont {Luetkens}}, \bibinfo {author} {\bibfnamefont
  {Z.}~\bibnamefont {Wang}}, \bibinfo {author} {\bibfnamefont {D.}~\bibnamefont
  {Yan}}, \bibinfo {author} {\bibfnamefont {H.}~\bibnamefont {Luo}},\ and\
  \bibinfo {author} {\bibfnamefont {Y.}~\bibnamefont {Shi}},\ }\bibfield
  {title} {\bibinfo {title} {Superconducting gap structure of the
  noncentrosymmetric topological superconductor candidate {HfRuP}},\ }\href
  {https://doi.org/10.3390/magnetochemistry9050135} {\bibfield  {journal}
  {\bibinfo  {journal} {Magnetochemistry}\ }\textbf {\bibinfo {volume} {9}},\
  \bibinfo {pages} {135} (\bibinfo {year} {2023})}\BibitemShut {NoStop}%
\bibitem [{\citenamefont {Shang}\ \emph
  {et~al.}(2022{\natexlab{b}})\citenamefont {Shang}, \citenamefont {Tay},
  \citenamefont {Su}, \citenamefont {Yuan},\ and\ \citenamefont
  {Shiroka}}]{Shang2022b}%
  \BibitemOpen
  \bibfield  {author} {\bibinfo {author} {\bibfnamefont {T.}~\bibnamefont
  {Shang}}, \bibinfo {author} {\bibfnamefont {D.}~\bibnamefont {Tay}}, \bibinfo
  {author} {\bibfnamefont {H.}~\bibnamefont {Su}}, \bibinfo {author}
  {\bibfnamefont {H.~Q.}\ \bibnamefont {Yuan}},\ and\ \bibinfo {author}
  {\bibfnamefont {T.}~\bibnamefont {Shiroka}},\ }\bibfield  {title} {\bibinfo
  {title} {Evidence of fully gapped superconductivity in {NbReSi}: A combined
  {\textmu}{SR} and {NMR} study},\ }\href
  {https://doi.org/10.1103/PhysRevB.105.144506} {\bibfield  {journal} {\bibinfo
   {journal} {Phys. Rev. B}\ }\textbf {\bibinfo {volume} {105}},\ \bibinfo
  {pages} {144506} (\bibinfo {year} {2022}{\natexlab{b}})}\BibitemShut
  {NoStop}%
\bibitem [{\citenamefont {Sajilesh}\ \emph {et~al.}(2024)\citenamefont
  {Sajilesh}, \citenamefont {Kushwaha}, \citenamefont {Samanta}, \citenamefont
  {Tula}, \citenamefont {Meena}, \citenamefont {Srivastava}, \citenamefont
  {Singh}, \citenamefont {Biswas}, \citenamefont {Kanigel}, \citenamefont
  {Hillier}, \citenamefont {Ghosh},\ and\ \citenamefont
  {Singh}}]{Sajilesh2024}%
  \BibitemOpen
  \bibfield  {author} {\bibinfo {author} {\bibfnamefont {P.~K.}\ \bibnamefont
  {Sajilesh}}, \bibinfo {author} {\bibfnamefont {R.~K.}\ \bibnamefont
  {Kushwaha}}, \bibinfo {author} {\bibfnamefont {D.}~\bibnamefont {Samanta}},
  \bibinfo {author} {\bibfnamefont {T.}~\bibnamefont {Tula}}, \bibinfo {author}
  {\bibfnamefont {P.~K.}\ \bibnamefont {Meena}}, \bibinfo {author}
  {\bibfnamefont {S.}~\bibnamefont {Srivastava}}, \bibinfo {author}
  {\bibfnamefont {D.}~\bibnamefont {Singh}}, \bibinfo {author} {\bibfnamefont
  {P.~K.}\ \bibnamefont {Biswas}}, \bibinfo {author} {\bibfnamefont
  {A.}~\bibnamefont {Kanigel}}, \bibinfo {author} {\bibfnamefont {A.~D.}\
  \bibnamefont {Hillier}}, \bibinfo {author} {\bibfnamefont {S.~K.}\
  \bibnamefont {Ghosh}},\ and\ \bibinfo {author} {\bibfnamefont {R.~P.}\
  \bibnamefont {Singh}},\ }\bibfield  {title} {\bibinfo {title}
  {Time‐reversal symmetry breaking superconductivity in {HfRhGe}: {A}
  noncentrosymmetric {W}eyl semimetal},\ }\href
  {https://doi.org/10.1002/adma.202415721} {\bibfield  {journal} {\bibinfo
  {journal} {Adv. Mater.}\ ,\ \bibinfo {pages} {2415721}} (\bibinfo {year}
  {2024})}\BibitemShut {NoStop}%
\bibitem [{\citenamefont {Suter}\ and\ \citenamefont
  {Wojek}(2012)}]{Suter2012}%
  \BibitemOpen
  \bibfield  {author} {\bibinfo {author} {\bibfnamefont {A.}~\bibnamefont
  {Suter}}\ and\ \bibinfo {author} {\bibfnamefont {B.~M.}\ \bibnamefont
  {Wojek}},\ }\bibfield  {title} {\bibinfo {title} {Musrfit: {A} free
  platform-independent framework for $\mu${SR} data analysis},\ }\href
  {https://doi.org/10.1016/j.phpro.2012.04.042} {\bibfield  {journal} {\bibinfo
   {journal} {Phys. Procedia}\ }\textbf {\bibinfo {volume} {30}},\ \bibinfo
  {pages} {69} (\bibinfo {year} {2012})}\BibitemShut {NoStop}%
\bibitem [{\citenamefont {Giannozzi}\ \emph {et~al.}(2009)\citenamefont
  {Giannozzi}, \citenamefont {Baroni}, \citenamefont {Bonini}, \citenamefont
  {Calandra}, \citenamefont {Car}, \citenamefont {Cavazzoni}, \citenamefont
  {Ceresoli}, \citenamefont {Chiarotti}, \citenamefont {Cococcioni},
  \citenamefont {Dabo}, \citenamefont {Corso}, \citenamefont {de~Gironcoli},
  \citenamefont {Fabris}, \citenamefont {Fratesi}, \citenamefont {Gebauer},
  \citenamefont {Gerstmann}, \citenamefont {Gougoussis}, \citenamefont
  {Kokalj}, \citenamefont {Lazzeri}, \citenamefont {Martin-Samos},
  \citenamefont {Marzari}, \citenamefont {Mauri}, \citenamefont {Mazzarello},
  \citenamefont {Paolini}, \citenamefont {Pasquarello}, \citenamefont
  {Paulatto}, \citenamefont {Sbraccia}, \citenamefont {Scandolo}, \citenamefont
  {Sclauzero}, \citenamefont {Seitsonen}, \citenamefont {Smogunov},
  \citenamefont {Umari},\ and\ \citenamefont {Wentzcovitch}}]{giannozzi2009}%
  \BibitemOpen
  \bibfield  {author} {\bibinfo {author} {\bibfnamefont {P.}~\bibnamefont
  {Giannozzi}}, \bibinfo {author} {\bibfnamefont {S.}~\bibnamefont {Baroni}},
  \bibinfo {author} {\bibfnamefont {N.}~\bibnamefont {Bonini}}, \bibinfo
  {author} {\bibfnamefont {M.}~\bibnamefont {Calandra}}, \bibinfo {author}
  {\bibfnamefont {R.}~\bibnamefont {Car}}, \bibinfo {author} {\bibfnamefont
  {C.}~\bibnamefont {Cavazzoni}}, \bibinfo {author} {\bibfnamefont
  {D.}~\bibnamefont {Ceresoli}}, \bibinfo {author} {\bibfnamefont {G.~L.}\
  \bibnamefont {Chiarotti}}, \bibinfo {author} {\bibfnamefont {M.}~\bibnamefont
  {Cococcioni}}, \bibinfo {author} {\bibfnamefont {I.}~\bibnamefont {Dabo}},
  \bibinfo {author} {\bibfnamefont {A.~D.}\ \bibnamefont {Corso}}, \bibinfo
  {author} {\bibfnamefont {S.}~\bibnamefont {de~Gironcoli}}, \bibinfo {author}
  {\bibfnamefont {S.}~\bibnamefont {Fabris}}, \bibinfo {author} {\bibfnamefont
  {G.}~\bibnamefont {Fratesi}}, \bibinfo {author} {\bibfnamefont
  {R.}~\bibnamefont {Gebauer}}, \bibinfo {author} {\bibfnamefont
  {U.}~\bibnamefont {Gerstmann}}, \bibinfo {author} {\bibfnamefont
  {C.}~\bibnamefont {Gougoussis}}, \bibinfo {author} {\bibfnamefont
  {A.}~\bibnamefont {Kokalj}}, \bibinfo {author} {\bibfnamefont
  {M.}~\bibnamefont {Lazzeri}}, \bibinfo {author} {\bibfnamefont
  {L.}~\bibnamefont {Martin-Samos}}, \bibinfo {author} {\bibfnamefont
  {N.}~\bibnamefont {Marzari}}, \bibinfo {author} {\bibfnamefont
  {F.}~\bibnamefont {Mauri}}, \bibinfo {author} {\bibfnamefont
  {R.}~\bibnamefont {Mazzarello}}, \bibinfo {author} {\bibfnamefont
  {S.}~\bibnamefont {Paolini}}, \bibinfo {author} {\bibfnamefont
  {A.}~\bibnamefont {Pasquarello}}, \bibinfo {author} {\bibfnamefont
  {L.}~\bibnamefont {Paulatto}}, \bibinfo {author} {\bibfnamefont
  {C.}~\bibnamefont {Sbraccia}}, \bibinfo {author} {\bibfnamefont
  {S.}~\bibnamefont {Scandolo}}, \bibinfo {author} {\bibfnamefont
  {G.}~\bibnamefont {Sclauzero}}, \bibinfo {author} {\bibfnamefont {A.~P.}\
  \bibnamefont {Seitsonen}}, \bibinfo {author} {\bibfnamefont {A.}~\bibnamefont
  {Smogunov}}, \bibinfo {author} {\bibfnamefont {P.}~\bibnamefont {Umari}},\
  and\ \bibinfo {author} {\bibfnamefont {R.~M.}\ \bibnamefont {Wentzcovitch}},\
  }\bibfield  {title} {\bibinfo {title} {{{QUANTUM} {ESPRESSO}: a modular and
  open-source software project for quantum simulations of materials}},\ }\href
  {https://doi.org/10.1088/0953-8984/21/39/395502} {\bibfield  {journal}
  {\bibinfo  {journal} {J. Phys.: Condens. Matter}\ }\textbf {\bibinfo {volume}
  {21}},\ \bibinfo {pages} {395502} (\bibinfo {year} {2009})}\BibitemShut
  {NoStop}%
\bibitem [{\citenamefont {Giannozzi}\ \emph {et~al.}(2017)\citenamefont
  {Giannozzi}, \citenamefont {Andreussi}, \citenamefont {Brumme}, \citenamefont
  {Bunau}, \citenamefont {Nardelli}, \citenamefont {Calandra}, \citenamefont
  {Car}, \citenamefont {Cavazzoni}, \citenamefont {Ceresoli}, \citenamefont
  {Cococcioni}, \citenamefont {Colonna}, \citenamefont {Carnimeo},
  \citenamefont {Corso}, \citenamefont {de~Gironcoli}, \citenamefont {Delugas},
  \citenamefont {DiStasio}, \citenamefont {Ferretti}, \citenamefont {Floris},
  \citenamefont {Fratesi}, \citenamefont {Fugallo}, \citenamefont {Gebauer},
  \citenamefont {Gerstmann}, \citenamefont {Giustino}, \citenamefont {Gorni},
  \citenamefont {Jia}, \citenamefont {Kawamura}, \citenamefont {Ko},
  \citenamefont {Kokalj}, \citenamefont {K{\"{u}}{\c{c}}{\"{u}}kbenli},
  \citenamefont {Lazzeri}, \citenamefont {Marsili}, \citenamefont {Marzari},
  \citenamefont {Mauri}, \citenamefont {Nguyen}, \citenamefont {Nguyen},
  \citenamefont {Otero-de-la Roza}, \citenamefont {Paulatto}, \citenamefont
  {Ponc{\'{e}}}, \citenamefont {Rocca}, \citenamefont {Sabatini}, \citenamefont
  {Santra}, \citenamefont {Schlipf}, \citenamefont {Seitsonen}, \citenamefont
  {Smogunov}, \citenamefont {Timrov}, \citenamefont {Thonhauser}, \citenamefont
  {Umari}, \citenamefont {Vast}, \citenamefont {Wu},\ and\ \citenamefont
  {Baroni}}]{giannozzi2017}%
  \BibitemOpen
  \bibfield  {author} {\bibinfo {author} {\bibfnamefont {P.}~\bibnamefont
  {Giannozzi}}, \bibinfo {author} {\bibfnamefont {O.}~\bibnamefont
  {Andreussi}}, \bibinfo {author} {\bibfnamefont {T.}~\bibnamefont {Brumme}},
  \bibinfo {author} {\bibfnamefont {O.}~\bibnamefont {Bunau}}, \bibinfo
  {author} {\bibfnamefont {M.~B.}\ \bibnamefont {Nardelli}}, \bibinfo {author}
  {\bibfnamefont {M.}~\bibnamefont {Calandra}}, \bibinfo {author}
  {\bibfnamefont {R.}~\bibnamefont {Car}}, \bibinfo {author} {\bibfnamefont
  {C.}~\bibnamefont {Cavazzoni}}, \bibinfo {author} {\bibfnamefont
  {D.}~\bibnamefont {Ceresoli}}, \bibinfo {author} {\bibfnamefont
  {M.}~\bibnamefont {Cococcioni}}, \bibinfo {author} {\bibfnamefont
  {N.}~\bibnamefont {Colonna}}, \bibinfo {author} {\bibfnamefont
  {I.}~\bibnamefont {Carnimeo}}, \bibinfo {author} {\bibfnamefont {A.~D.}\
  \bibnamefont {Corso}}, \bibinfo {author} {\bibfnamefont {S.}~\bibnamefont
  {de~Gironcoli}}, \bibinfo {author} {\bibfnamefont {P.}~\bibnamefont
  {Delugas}}, \bibinfo {author} {\bibfnamefont {R.~A.}\ \bibnamefont
  {DiStasio}}, \bibinfo {author} {\bibfnamefont {A.}~\bibnamefont {Ferretti}},
  \bibinfo {author} {\bibfnamefont {A.}~\bibnamefont {Floris}}, \bibinfo
  {author} {\bibfnamefont {G.}~\bibnamefont {Fratesi}}, \bibinfo {author}
  {\bibfnamefont {G.}~\bibnamefont {Fugallo}}, \bibinfo {author} {\bibfnamefont
  {R.}~\bibnamefont {Gebauer}}, \bibinfo {author} {\bibfnamefont
  {U.}~\bibnamefont {Gerstmann}}, \bibinfo {author} {\bibfnamefont
  {F.}~\bibnamefont {Giustino}}, \bibinfo {author} {\bibfnamefont
  {T.}~\bibnamefont {Gorni}}, \bibinfo {author} {\bibfnamefont
  {J.}~\bibnamefont {Jia}}, \bibinfo {author} {\bibfnamefont {M.}~\bibnamefont
  {Kawamura}}, \bibinfo {author} {\bibfnamefont {H.-Y.}\ \bibnamefont {Ko}},
  \bibinfo {author} {\bibfnamefont {A.}~\bibnamefont {Kokalj}}, \bibinfo
  {author} {\bibfnamefont {E.}~\bibnamefont {K{\"{u}}{\c{c}}{\"{u}}kbenli}},
  \bibinfo {author} {\bibfnamefont {M.}~\bibnamefont {Lazzeri}}, \bibinfo
  {author} {\bibfnamefont {M.}~\bibnamefont {Marsili}}, \bibinfo {author}
  {\bibfnamefont {N.}~\bibnamefont {Marzari}}, \bibinfo {author} {\bibfnamefont
  {F.}~\bibnamefont {Mauri}}, \bibinfo {author} {\bibfnamefont {N.~L.}\
  \bibnamefont {Nguyen}}, \bibinfo {author} {\bibfnamefont {H.-V.}\
  \bibnamefont {Nguyen}}, \bibinfo {author} {\bibfnamefont {A.}~\bibnamefont
  {Otero-de-la Roza}}, \bibinfo {author} {\bibfnamefont {L.}~\bibnamefont
  {Paulatto}}, \bibinfo {author} {\bibfnamefont {S.}~\bibnamefont
  {Ponc{\'{e}}}}, \bibinfo {author} {\bibfnamefont {D.}~\bibnamefont {Rocca}},
  \bibinfo {author} {\bibfnamefont {R.}~\bibnamefont {Sabatini}}, \bibinfo
  {author} {\bibfnamefont {B.}~\bibnamefont {Santra}}, \bibinfo {author}
  {\bibfnamefont {M.}~\bibnamefont {Schlipf}}, \bibinfo {author} {\bibfnamefont
  {A.~P.}\ \bibnamefont {Seitsonen}}, \bibinfo {author} {\bibfnamefont
  {A.}~\bibnamefont {Smogunov}}, \bibinfo {author} {\bibfnamefont
  {I.}~\bibnamefont {Timrov}}, \bibinfo {author} {\bibfnamefont
  {T.}~\bibnamefont {Thonhauser}}, \bibinfo {author} {\bibfnamefont
  {P.}~\bibnamefont {Umari}}, \bibinfo {author} {\bibfnamefont
  {N.}~\bibnamefont {Vast}}, \bibinfo {author} {\bibfnamefont {X.}~\bibnamefont
  {Wu}},\ and\ \bibinfo {author} {\bibfnamefont {S.}~\bibnamefont {Baroni}},\
  }\bibfield  {title} {\bibinfo {title} {{Advanced capabilities for materials
  modelling with Quantum {ESPRESSO}}},\ }\href
  {https://doi.org/10.1088/1361-648x/aa8f79} {\bibfield  {journal} {\bibinfo
  {journal} {J. Phys.: Condens. Matter}\ }\textbf {\bibinfo {volume} {29}},\
  \bibinfo {pages} {465901} (\bibinfo {year} {2017})}\BibitemShut {NoStop}%
\bibitem [{\citenamefont {Perdew}\ \emph {et~al.}(1996)\citenamefont {Perdew},
  \citenamefont {Burke},\ and\ \citenamefont {Ernzerhof}}]{Perdew1996iq}%
  \BibitemOpen
  \bibfield  {author} {\bibinfo {author} {\bibfnamefont {J.~P.}\ \bibnamefont
  {Perdew}}, \bibinfo {author} {\bibfnamefont {K.}~\bibnamefont {Burke}},\ and\
  \bibinfo {author} {\bibfnamefont {M.}~\bibnamefont {Ernzerhof}},\ }\bibfield
  {title} {\bibinfo {title} {Generalized gradient approximation made simple},\
  }\href {https://doi.org/10.1103/PhysRevLett.77.3865} {\bibfield  {journal}
  {\bibinfo  {journal} {Phys. Rev. Lett.}\ }\textbf {\bibinfo {volume} {77}},\
  \bibinfo {pages} {3865} (\bibinfo {year} {1996})}\BibitemShut {NoStop}%
\bibitem [{\citenamefont {Bl\"ochl}(1994)}]{Blochl1994zz}%
  \BibitemOpen
  \bibfield  {author} {\bibinfo {author} {\bibfnamefont {P.~E.}\ \bibnamefont
  {Bl\"ochl}},\ }\bibfield  {title} {\bibinfo {title} {Projector augmented-wave
  method},\ }\href {https://doi.org/10.1103/PhysRevB.50.17953} {\bibfield
  {journal} {\bibinfo  {journal} {Phys. Rev. B}\ }\textbf {\bibinfo {volume}
  {50}},\ \bibinfo {pages} {17953} (\bibinfo {year} {1994})}\BibitemShut
  {NoStop}%
\bibitem [{\citenamefont {Zhu}\ \emph {et~al.}(2008)\citenamefont {Zhu},
  \citenamefont {Yang}, \citenamefont {Fang}, \citenamefont {Mu},\ and\
  \citenamefont {Wen}}]{Zhu2008}%
  \BibitemOpen
  \bibfield  {author} {\bibinfo {author} {\bibfnamefont {X.}~\bibnamefont
  {Zhu}}, \bibinfo {author} {\bibfnamefont {H.}~\bibnamefont {Yang}}, \bibinfo
  {author} {\bibfnamefont {L.}~\bibnamefont {Fang}}, \bibinfo {author}
  {\bibfnamefont {G.}~\bibnamefont {Mu}},\ and\ \bibinfo {author}
  {\bibfnamefont {H.-H.}\ \bibnamefont {Wen}},\ }\bibfield  {title} {\bibinfo
  {title} {Upper critical field, {H}all effect and magnetoresistance in the
  iron-based layered superconductor {LaFeAsO}$_{0.9}${F}$_{0.1-\delta}$},\
  }\href {https://doi.org/10.1088/0953-2048/21/10/105001} {\bibfield  {journal}
  {\bibinfo  {journal} {Supercond. Sci. Technol.}\ }\textbf {\bibinfo {volume}
  {21}},\ \bibinfo {pages} {105001} (\bibinfo {year} {2008})}\BibitemShut
  {NoStop}%
\bibitem [{\citenamefont {Werthamer}\ \emph {et~al.}(1966)\citenamefont
  {Werthamer}, \citenamefont {Helfand},\ and\ \citenamefont
  {Hohenberg}}]{Werthamer1966}%
  \BibitemOpen
  \bibfield  {author} {\bibinfo {author} {\bibfnamefont {N.~R.}\ \bibnamefont
  {Werthamer}}, \bibinfo {author} {\bibfnamefont {E.}~\bibnamefont {Helfand}},\
  and\ \bibinfo {author} {\bibfnamefont {P.~C.}\ \bibnamefont {Hohenberg}},\
  }\bibfield  {title} {\bibinfo {title} {Temperature and purity dependence of
  the superconducting critical field, ${{H}}_{c2}$. {III}. {E}lectron spin and
  spin-orbit effects},\ }\href {https://doi.org/10.1103/PhysRev.147.295}
  {\bibfield  {journal} {\bibinfo  {journal} {Phys. Rev.}\ }\textbf {\bibinfo
  {volume} {147}},\ \bibinfo {pages} {295} (\bibinfo {year}
  {1966})}\BibitemShut {NoStop}%
\bibitem [{\citenamefont {Gurevich}(2011)}]{Gurevich2011}%
  \BibitemOpen
  \bibfield  {author} {\bibinfo {author} {\bibfnamefont {A.}~\bibnamefont
  {Gurevich}},\ }\bibfield  {title} {\bibinfo {title} {Iron-based
  superconductors at high magnetic fields},\ }\href
  {https://doi.org/10.1088/0034-4885/74/12/124501} {\bibfield  {journal}
  {\bibinfo  {journal} {Rep. Prog. Phys}\ }\textbf {\bibinfo {volume} {74}},\
  \bibinfo {pages} {124501} (\bibinfo {year} {2011})},\ \bibinfo {note} {and
  references therein}\BibitemShut {NoStop}%
\bibitem [{\citenamefont {Nakajima}\ \emph {et~al.}(2012)\citenamefont
  {Nakajima}, \citenamefont {Hidaka}, \citenamefont {Nakagawa}, \citenamefont
  {Tamegai}, \citenamefont {Nishizaki}, \citenamefont {Sasaki},\ and\
  \citenamefont {Kobayashi}}]{Nakajima2012}%
  \BibitemOpen
  \bibfield  {author} {\bibinfo {author} {\bibfnamefont {Y.}~\bibnamefont
  {Nakajima}}, \bibinfo {author} {\bibfnamefont {H.}~\bibnamefont {Hidaka}},
  \bibinfo {author} {\bibfnamefont {T.}~\bibnamefont {Nakagawa}}, \bibinfo
  {author} {\bibfnamefont {T.}~\bibnamefont {Tamegai}}, \bibinfo {author}
  {\bibfnamefont {T.}~\bibnamefont {Nishizaki}}, \bibinfo {author}
  {\bibfnamefont {T.}~\bibnamefont {Sasaki}},\ and\ \bibinfo {author}
  {\bibfnamefont {N.}~\bibnamefont {Kobayashi}},\ }\bibfield  {title} {\bibinfo
  {title} {Two-band superconductivity featuring different anisotropies in the
  ternary iron silicide {Lu}$_2${Fe}$_3${Si}$_5$},\ }\href
  {https://doi.org/10.1103/PhysRevB.85.174524} {\bibfield  {journal} {\bibinfo
  {journal} {Phys. Rev. B}\ }\textbf {\bibinfo {volume} {85}},\ \bibinfo
  {pages} {174524} (\bibinfo {year} {2012})}\BibitemShut {NoStop}%
\bibitem [{\citenamefont {M\"{u}ller}\ \emph {et~al.}(2001)\citenamefont
  {M\"{u}ller}, \citenamefont {Fuchs}, \citenamefont {Handstein}, \citenamefont
  {Nenkov}, \citenamefont {Narozhnyi},\ and\ \citenamefont
  {Eckert}}]{Muller2001}%
  \BibitemOpen
  \bibfield  {author} {\bibinfo {author} {\bibfnamefont {K.-H.}\ \bibnamefont
  {M\"{u}ller}}, \bibinfo {author} {\bibfnamefont {G.}~\bibnamefont {Fuchs}},
  \bibinfo {author} {\bibfnamefont {A.}~\bibnamefont {Handstein}}, \bibinfo
  {author} {\bibfnamefont {K.}~\bibnamefont {Nenkov}}, \bibinfo {author}
  {\bibfnamefont {V.~N.}\ \bibnamefont {Narozhnyi}},\ and\ \bibinfo {author}
  {\bibfnamefont {D.}~\bibnamefont {Eckert}},\ }\bibfield  {title} {\bibinfo
  {title} {The upper critical field in superconducting {Mg}{B}$_2$},\ }\href
  {https://doi.org/10.1016/S0925-8388(01)01197-5} {\bibfield  {journal}
  {\bibinfo  {journal} {J. Alloys Compd.}\ }\textbf {\bibinfo {volume} {322}},\
  \bibinfo {pages} {L10} (\bibinfo {year} {2001})}\BibitemShut {NoStop}%
\bibitem [{\citenamefont {Gurevich}\ \emph {et~al.}(2004)\citenamefont
  {Gurevich}, \citenamefont {Patnaik}, \citenamefont {Braccini}, \citenamefont
  {Kim}, \citenamefont {Mielke}, \citenamefont {Song}, \citenamefont {Cooley},
  \citenamefont {Bu}, \citenamefont {Kim}, \citenamefont {Choi}, \citenamefont
  {Belenky}, \citenamefont {Giencke}, \citenamefont {Lee}, \citenamefont
  {Tian}, \citenamefont {Pan}, \citenamefont {Siri}, \citenamefont {Hellstrom},
  \citenamefont {Eom},\ and\ \citenamefont {Larbalestier}}]{Gurevich2004}%
  \BibitemOpen
  \bibfield  {author} {\bibinfo {author} {\bibfnamefont {A.}~\bibnamefont
  {Gurevich}}, \bibinfo {author} {\bibfnamefont {S.}~\bibnamefont {Patnaik}},
  \bibinfo {author} {\bibfnamefont {V.}~\bibnamefont {Braccini}}, \bibinfo
  {author} {\bibfnamefont {K.~H.}\ \bibnamefont {Kim}}, \bibinfo {author}
  {\bibfnamefont {C.}~\bibnamefont {Mielke}}, \bibinfo {author} {\bibfnamefont
  {X.}~\bibnamefont {Song}}, \bibinfo {author} {\bibfnamefont {L.~D.}\
  \bibnamefont {Cooley}}, \bibinfo {author} {\bibfnamefont {S.~D.}\
  \bibnamefont {Bu}}, \bibinfo {author} {\bibfnamefont {D.~M.}\ \bibnamefont
  {Kim}}, \bibinfo {author} {\bibfnamefont {J.~H.}\ \bibnamefont {Choi}},
  \bibinfo {author} {\bibfnamefont {L.~J.}\ \bibnamefont {Belenky}}, \bibinfo
  {author} {\bibfnamefont {J.}~\bibnamefont {Giencke}}, \bibinfo {author}
  {\bibfnamefont {M.~K.}\ \bibnamefont {Lee}}, \bibinfo {author} {\bibfnamefont
  {W.}~\bibnamefont {Tian}}, \bibinfo {author} {\bibfnamefont {X.~Q.}\
  \bibnamefont {Pan}}, \bibinfo {author} {\bibfnamefont {A.}~\bibnamefont
  {Siri}}, \bibinfo {author} {\bibfnamefont {E.~E.}\ \bibnamefont {Hellstrom}},
  \bibinfo {author} {\bibfnamefont {C.~B.}\ \bibnamefont {Eom}},\ and\ \bibinfo
  {author} {\bibfnamefont {D.~C.}\ \bibnamefont {Larbalestier}},\ }\bibfield
  {title} {\bibinfo {title} {Very high upper critical fields in {Mg}{B}$_2$
  produced by selective tuning of impurity scattering},\ }\href
  {https://doi.org/10.1088/0953-2048/17/2/008} {\bibfield  {journal} {\bibinfo
  {journal} {Supercond. Sci. Technol.}\ }\textbf {\bibinfo {volume} {17}},\
  \bibinfo {pages} {278} (\bibinfo {year} {2004})}\BibitemShut {NoStop}%
\bibitem [{\citenamefont {Shang}\ \emph {et~al.}(2021)\citenamefont {Shang},
  \citenamefont {Xie}, \citenamefont {Zhao}, \citenamefont {Chen},
  \citenamefont {Gawryluk}, \citenamefont {Medarde}, \citenamefont {Shi},
  \citenamefont {Yuan}, \citenamefont {Pomjakushina},\ and\ \citenamefont
  {Shiroka}}]{Shang2021}%
  \BibitemOpen
  \bibfield  {author} {\bibinfo {author} {\bibfnamefont {T.}~\bibnamefont
  {Shang}}, \bibinfo {author} {\bibfnamefont {W.}~\bibnamefont {Xie}}, \bibinfo
  {author} {\bibfnamefont {J.~Z.}\ \bibnamefont {Zhao}}, \bibinfo {author}
  {\bibfnamefont {Y.}~\bibnamefont {Chen}}, \bibinfo {author} {\bibfnamefont
  {D.~J.}\ \bibnamefont {Gawryluk}}, \bibinfo {author} {\bibfnamefont
  {M.}~\bibnamefont {Medarde}}, \bibinfo {author} {\bibfnamefont
  {M.}~\bibnamefont {Shi}}, \bibinfo {author} {\bibfnamefont {H.~Q.}\
  \bibnamefont {Yuan}}, \bibinfo {author} {\bibfnamefont {E.}~\bibnamefont
  {Pomjakushina}},\ and\ \bibinfo {author} {\bibfnamefont {T.}~\bibnamefont
  {Shiroka}},\ }\bibfield  {title} {\bibinfo {title} {Multigap
  superconductivity in centrosymmetric and noncentrosymmetric rhenium-boron
  superconductors},\ }\href {https://doi.org/10.1103/PhysRevB.103.184517}
  {\bibfield  {journal} {\bibinfo  {journal} {Phys. Rev. B}\ }\textbf {\bibinfo
  {volume} {103}},\ \bibinfo {pages} {184517} (\bibinfo {year}
  {2021})}\BibitemShut {NoStop}%
\bibitem [{\citenamefont {Maisuradze}\ \emph {et~al.}(2009)\citenamefont
  {Maisuradze}, \citenamefont {Khasanov}, \citenamefont {Shengelaya},\ and\
  \citenamefont {Keller}}]{Maisuradze2009}%
  \BibitemOpen
  \bibfield  {author} {\bibinfo {author} {\bibfnamefont {A.}~\bibnamefont
  {Maisuradze}}, \bibinfo {author} {\bibfnamefont {R.}~\bibnamefont
  {Khasanov}}, \bibinfo {author} {\bibfnamefont {A.}~\bibnamefont
  {Shengelaya}},\ and\ \bibinfo {author} {\bibfnamefont {H.}~\bibnamefont
  {Keller}},\ }\bibfield  {title} {\bibinfo {title} {Comparison of different
  methods for analyzing $\mu${SR} line shapes in the vortex state of type-{II}
  superconductors},\ }\href {https://doi.org/doi:10.1088/0953-8984/21/7/075701}
  {\bibfield  {journal} {\bibinfo  {journal} {J. Phys.: Condens. Mat.}\
  }\textbf {\bibinfo {volume} {21}},\ \bibinfo {pages} {075701} (\bibinfo
  {year} {2009})},\ \bibinfo {note} {and references therein}\BibitemShut
  {NoStop}%
\bibitem [{\citenamefont {Barford}\ and\ \citenamefont
  {Gunn}(1988)}]{Barford1988}%
  \BibitemOpen
  \bibfield  {author} {\bibinfo {author} {\bibfnamefont {W.}~\bibnamefont
  {Barford}}\ and\ \bibinfo {author} {\bibfnamefont {J.~M.~F.}\ \bibnamefont
  {Gunn}},\ }\bibfield  {title} {\bibinfo {title} {The theory of the
  measurement of the {L}ondon penetration depth in uniaxial type-{II}
  superconductors by muon spin rotation},\ }\href
  {https://doi.org/10.1016/0921-4534(88)90014-7} {\bibfield  {journal}
  {\bibinfo  {journal} {Physica C}\ }\textbf {\bibinfo {volume} {156}},\
  \bibinfo {pages} {515} (\bibinfo {year} {1988})}\BibitemShut {NoStop}%
\bibitem [{\citenamefont {Brandt}(2003)}]{Brandt2003}%
  \BibitemOpen
  \bibfield  {author} {\bibinfo {author} {\bibfnamefont {E.~H.}\ \bibnamefont
  {Brandt}},\ }\bibfield  {title} {\bibinfo {title} {Properties of the ideal
  {G}inzburg-{L}andau vortex lattice},\ }\href
  {https://doi.org/10.1103/PhysRevB.68.054506} {\bibfield  {journal} {\bibinfo
  {journal} {Phys. Rev. B}\ }\textbf {\bibinfo {volume} {68}},\ \bibinfo
  {pages} {054506} (\bibinfo {year} {2003})}\BibitemShut {NoStop}%
\bibitem [{\citenamefont {Serventi}\ \emph {et~al.}(2004)\citenamefont
  {Serventi}, \citenamefont {Allodi}, \citenamefont {De~Renzi}, \citenamefont
  {Guidi}, \citenamefont {Roman\`o}, \citenamefont {Manfrinetti}, \citenamefont
  {Palenzona}, \citenamefont {Niedermayer}, \citenamefont {Amato},\ and\
  \citenamefont {Baines}}]{Serventi2004}%
  \BibitemOpen
  \bibfield  {author} {\bibinfo {author} {\bibfnamefont {S.}~\bibnamefont
  {Serventi}}, \bibinfo {author} {\bibfnamefont {G.}~\bibnamefont {Allodi}},
  \bibinfo {author} {\bibfnamefont {R.}~\bibnamefont {De~Renzi}}, \bibinfo
  {author} {\bibfnamefont {G.}~\bibnamefont {Guidi}}, \bibinfo {author}
  {\bibfnamefont {L.}~\bibnamefont {Roman\`o}}, \bibinfo {author}
  {\bibfnamefont {P.}~\bibnamefont {Manfrinetti}}, \bibinfo {author}
  {\bibfnamefont {A.}~\bibnamefont {Palenzona}}, \bibinfo {author}
  {\bibfnamefont {C.}~\bibnamefont {Niedermayer}}, \bibinfo {author}
  {\bibfnamefont {A.}~\bibnamefont {Amato}},\ and\ \bibinfo {author}
  {\bibfnamefont {C.}~\bibnamefont {Baines}},\ }\bibfield  {title} {\bibinfo
  {title} {{E}ffect of two gaps on the flux-lattice internal field
  distribution: {E}vidence of two length scales in {Mg}$_{1-x}${Al}$_x${B}$_2$
  from $\mu${SR}},\ }\href {https://doi.org/10.1103/PhysRevLett.93.217003}
  {\bibfield  {journal} {\bibinfo  {journal} {Phys. Rev. Lett.}\ }\textbf
  {\bibinfo {volume} {93}},\ \bibinfo {pages} {217003} (\bibinfo {year}
  {2004})}\BibitemShut {NoStop}%
\bibitem [{\citenamefont {Khasanov}\ \emph {et~al.}(2014)\citenamefont
  {Khasanov}, \citenamefont {Amato}, \citenamefont {Biswas}, \citenamefont
  {Luetkens}, \citenamefont {Zhigadlo},\ and\ \citenamefont
  {Batlogg}}]{Khasanov2014}%
  \BibitemOpen
  \bibfield  {author} {\bibinfo {author} {\bibfnamefont {R.}~\bibnamefont
  {Khasanov}}, \bibinfo {author} {\bibfnamefont {A.}~\bibnamefont {Amato}},
  \bibinfo {author} {\bibfnamefont {P.~K.}\ \bibnamefont {Biswas}}, \bibinfo
  {author} {\bibfnamefont {H.}~\bibnamefont {Luetkens}}, \bibinfo {author}
  {\bibfnamefont {N.~D.}\ \bibnamefont {Zhigadlo}},\ and\ \bibinfo {author}
  {\bibfnamefont {B.}~\bibnamefont {Batlogg}},\ }\bibfield  {title} {\bibinfo
  {title} {{Sr}{Pt}$_3${P}: {A} two-band single-gap superconductor},\ }\href
  {https://doi.org/10.1103/PhysRevB.90.140507} {\bibfield  {journal} {\bibinfo
  {journal} {Phys. Rev. B}\ }\textbf {\bibinfo {volume} {90}},\ \bibinfo
  {pages} {140507(R)} (\bibinfo {year} {2014})}\BibitemShut {NoStop}%
\bibitem [{\citenamefont {Khasanov}\ \emph {et~al.}(2020)\citenamefont
  {Khasanov}, \citenamefont {Gupta}, \citenamefont {Das}, \citenamefont
  {Leithe-Jasper},\ and\ \citenamefont {Svanidze}}]{Khasanov2020}%
  \BibitemOpen
  \bibfield  {author} {\bibinfo {author} {\bibfnamefont {R.}~\bibnamefont
  {Khasanov}}, \bibinfo {author} {\bibfnamefont {R.}~\bibnamefont {Gupta}},
  \bibinfo {author} {\bibfnamefont {D.}~\bibnamefont {Das}}, \bibinfo {author}
  {\bibfnamefont {A.}~\bibnamefont {Leithe-Jasper}},\ and\ \bibinfo {author}
  {\bibfnamefont {E.}~\bibnamefont {Svanidze}},\ }\bibfield  {title} {\bibinfo
  {title} {Single-gap versus two-gap scenario: Specific heat and thermodynamic
  critical field of the noncentrosymmetric superconductor {BeAu}},\ }\href
  {https://doi.org/10.1103/PhysRevB.102.014514} {\bibfield  {journal} {\bibinfo
   {journal} {Phys. Rev. B}\ }\textbf {\bibinfo {volume} {102}},\ \bibinfo
  {pages} {014514} (\bibinfo {year} {2020})}\BibitemShut {NoStop}%
\bibitem [{\citenamefont {Shang}\ \emph {et~al.}(2019)\citenamefont {Shang},
  \citenamefont {Amon}, \citenamefont {Kasinathan}, \citenamefont {Xie},
  \citenamefont {Bobnar}, \citenamefont {Chen}, \citenamefont {Wang},
  \citenamefont {Shi}, \citenamefont {Medarde}, \citenamefont {Yuan},\ and\
  \citenamefont {Shiroka}}]{Shang2019c}%
  \BibitemOpen
  \bibfield  {author} {\bibinfo {author} {\bibfnamefont {T.}~\bibnamefont
  {Shang}}, \bibinfo {author} {\bibfnamefont {A.}~\bibnamefont {Amon}},
  \bibinfo {author} {\bibfnamefont {D.}~\bibnamefont {Kasinathan}}, \bibinfo
  {author} {\bibfnamefont {W.}~\bibnamefont {Xie}}, \bibinfo {author}
  {\bibfnamefont {M.}~\bibnamefont {Bobnar}}, \bibinfo {author} {\bibfnamefont
  {Y.}~\bibnamefont {Chen}}, \bibinfo {author} {\bibfnamefont {A.}~\bibnamefont
  {Wang}}, \bibinfo {author} {\bibfnamefont {M.}~\bibnamefont {Shi}}, \bibinfo
  {author} {\bibfnamefont {M.}~\bibnamefont {Medarde}}, \bibinfo {author}
  {\bibfnamefont {H.~Q.}\ \bibnamefont {Yuan}},\ and\ \bibinfo {author}
  {\bibfnamefont {T.}~\bibnamefont {Shiroka}},\ }\bibfield  {title} {\bibinfo
  {title} {Enhanced ${T}_c$ and multiband superconductivity in the fully-gapped
  {ReBe}$_{22}$ superconductor},\ }\href
  {https://doi.org/10.1088/1367-2630/ab307b} {\bibfield  {journal} {\bibinfo
  {journal} {New J. Phys.}\ }\textbf {\bibinfo {volume} {21}},\ \bibinfo
  {pages} {073034} (\bibinfo {year} {2019})}\BibitemShut {NoStop}%
\bibitem [{\citenamefont {Shang}\ \emph
  {et~al.}(2020{\natexlab{c}})\citenamefont {Shang}, \citenamefont {Xie},
  \citenamefont {Gawryluk}, \citenamefont {Khasanov}, \citenamefont {Zhao},
  \citenamefont {Medarde}, \citenamefont {Shi}, \citenamefont {Yuan},
  \citenamefont {Pomjakushina},\ and\ \citenamefont {Shiroka}}]{Shang2020MoPB}%
  \BibitemOpen
  \bibfield  {author} {\bibinfo {author} {\bibfnamefont {T.}~\bibnamefont
  {Shang}}, \bibinfo {author} {\bibfnamefont {W.}~\bibnamefont {Xie}}, \bibinfo
  {author} {\bibfnamefont {D.~J.}\ \bibnamefont {Gawryluk}}, \bibinfo {author}
  {\bibfnamefont {R.}~\bibnamefont {Khasanov}}, \bibinfo {author}
  {\bibfnamefont {J.~Z.}\ \bibnamefont {Zhao}}, \bibinfo {author}
  {\bibfnamefont {M.}~\bibnamefont {Medarde}}, \bibinfo {author} {\bibfnamefont
  {M.}~\bibnamefont {Shi}}, \bibinfo {author} {\bibfnamefont {H.~Q.}\
  \bibnamefont {Yuan}}, \bibinfo {author} {\bibfnamefont {E.}~\bibnamefont
  {Pomjakushina}},\ and\ \bibinfo {author} {\bibfnamefont {T.}~\bibnamefont
  {Shiroka}},\ }\bibfield  {title} {\bibinfo {title} {Multigap
  superconductivity in the {Mo}$_5${PB}$_2$ boron{\textendash}phosphorus
  compound},\ }\href {https://doi.org/10.1088/1367-2630/abac3b} {\bibfield
  {journal} {\bibinfo  {journal} {New J. Phys.}\ }\textbf {\bibinfo {volume}
  {22}},\ \bibinfo {pages} {093016} (\bibinfo {year}
  {2020}{\natexlab{c}})}\BibitemShut {NoStop}%
\end{thebibliography}%
%\end{footnotesize}

\end{document}